\documentclass[aps,prd,onecolumn,nofootinbib,superscriptaddress]{revtex4}
\pdfoutput=1
\usepackage{graphicx}%Include figure files
\usepackage{subfigure}
\usepackage{amsmath}
\usepackage{amsfonts}
\usepackage{amssymb,ulem}
\usepackage{mathrsfs}
\usepackage{color}%
\usepackage{dcolumn}% Align table columns on decimal pointl

\usepackage{MnSymbol,wasysym}
\usepackage{braket}
\usepackage{eurosym}
\usepackage{calrsfs}
\usepackage{multirow}
\usepackage{float}

\newcommand{\RNum}[1]{\uppercase\expandafter{\romannumeral #1\relax}}
\usepackage[colorlinks=true,linkcolor=blue,urlcolor=blue,citecolor=red]{hyperref}
\usepackage[title]{appendix}
\begin{document}
\baselineskip=0.5 cm

\begin{flushright}
CTPU-PTC-26-03
\end{flushright}

\title{Precessions and parameter constraints from quasiperiodic oscillations in a rotating charged black hole}

\author{R. H. Ali}
\email{hasnainali408@yzu.edu.cn}
\affiliation{Center for Gravitation and Cosmology, College of Physical Science and Technology, Yangzhou University, Yangzhou, 225009, China}

\author{Meng-He Wu}
\email{mhwu@njtc.edu.cn}
\affiliation{School of Physics and Electronic Information Engineering, Neijiang Normal University, Neijiang 641112, China}

\author{Hong Guo}
\email{guohong@ibs.re.kr}
\affiliation{Particle Theory and Cosmology Group, Center for Theoretical Physics of the Universe,
Institute for Basic Science (IBS), Daejeon, 34126, Republic of Korea}

\author{Xiao-Mei Kuang}
\email{xmeikuang@yzu.edu.cn}
\affiliation{Center for Gravitation and Cosmology, College of Physical Science and Technology, Yangzhou University, Yangzhou, 225009, China}

\begin{abstract}
\baselineskip=0.5 cm
We investigate quasi-periodic oscillations (QPOs) as a diagnostic tool for probing frame-dragging effects and accretion disk physics in the spacetime of a rotating regular magnetic black hole (BH). Specifically, we analyze the precession of bound orbits and the epicyclic oscillations of test-particles under small perturbations in the equatorial plane. We demonstrate how the BH’s nonminimal coupling parameter ($\lambda/M^4$) and dimensionless magnetic charge ($Q/M$) significantly influence the three fundamental epicyclic frequencies. By applying the relativistic precession model and employing Markov Chain Monte Carlo simulations (MCMC), we constrain the BH’s characteristic parameters, including mass, spin, magnetic charge, and nonminimal coupling, with the use of observational QPO data from five X-ray binaries: $GRO J1655-40$, $XTE J1859+226$, $H1743-322$, $XTE J1550-564$, and $GRS 1915+105$. Furthermore, we examine the Lense-Thirring, geodetic, and general spin precession frequencies of a test gyroscope attached to a stationary observer around the current BH. Our theoretical results indicate that the regular charged BH suppresses these precession frequencies compared with the Kerr BH case.
\end{abstract}

\maketitle
\newpage
\tableofcontents

\section{Introduction}
Einstein’s general relativity (GR) is a cornerstone of modern theoretical physics, corroborated by a wide range of observational verifications. Recently, black holes (BHs) have moved to the forefront of modern astrophysics due to groundbreaking observational progress. These include the precise monitoring of stellar orbits near the Galactic Center~\cite{Ghez:2008ms,abuter2020detection}, the direct imaging of supermassive BH shadows in $M87^{*}$ and $SgrA^{*}$ by the Event Horizon Telescope~\cite{akiyama2019first,akiyama2022first}, and the historic detection of gravitational waves (GWs)~\cite{LIGOScientific:2016aoc, LIGOScientific:2018mvr,LIGOScientific:2020ibl}. These milestones are in remarkable consensus with the Kerr BH paradigm. Upcoming facilities, such as the Next Generation Very Large Array~\cite{di2019next}, the Thirty Meter Telescope~\cite{TMTInternationalScienceDevelopmentTeamsTMTScienceAdvisoryCommittee:2015pvw}, and space-based interferometers such as LISA~\cite{LISA:2017pwj}, Taiji~\cite{Hu:2017mde}, and TianQin~\cite{TianQin:2015yph}, are expected to provide even more salient findings in the strong-gravity regime.

Despite these successes, singularities are an inherent prediction of GR, signaling the breakdown of the classical theory in the ultraviolet regime. For example, there are numerous theoretical scenarios where a visible singularity may arise~\cite{Joshi:2011zm}, which demonstrates a central naked singularity forming as the equilibrium end state of gravitational collapse for a general matter cloud. In particular, the Kerr spacetime solution describes a singularity concealed behind the event horizon. It is widely anticipated that a complete theory of quantum gravity will ultimately resolve these singularities. Before such a theory is established, regular BHs, defined by a non-singular core, have been proposed to explore semiclassical and quantum aspects of BHs. Bardeen’s first static regular BH~\cite{bardeen1968proceedings}, later interpreted as Einstein gravity coupled to a nonlinear magnetic monopole source~\cite{Ayon-Beato:2000mjt}, provided a physical basis for such geometries. Subsequently, a variety of regular BH geometries have been proposed, illustrating how non-singular interiors can be embedded in classical gravity and examined across their causal structure, thermodynamics, and observational signatures. Among them, regular BHs are generally categorized into two types: semiclassical models with exotic matter sources~\cite{Dymnikova:1992ux,Nicolini:2005vd,Balakin:2016mnn}, and quantum-motivated models incorporating quantum corrections~\cite{Roupas:2022gee,Borde:1996df,Bonanno:2000ep,Gambini:2008dy,Perez:2017cmj,Brahma:2020eos}. These geometries serve as a vital bridge between GR and quantum gravity theories, see the reviews~\cite{Torres:2022twv, Lan:2023cvz}.

To explore these regular geometries beyond standard GR, nonminimal coupling theories have been developed, where gravitational fields interact with gauge fields through curvature tensors. These frameworks are classified into five main types~\cite{Kleihaus:2002ee, Azam:2017izk}: scalar-curvature, nonminimal Einstein-Maxwell, Einstein-Yang-Mills (EYM), Einstein-Yang-Mills-Higgs, and nonminimal Einstein-Maxwell-Axion models~\cite{Goenner:2014mka, Hehl:1999bt, Mueller-Hoissen:1987nvb, Balakin:2006gv, Balakin:2015gpq}. The ongoing analysis specifically focuses on the nonminimally coupled Einstein-Yang-Mills theory, with spacetime curvature directly interacting with non-Abelian gauge fields~\cite{Balakin:2015gpq}, in which an analytical regular spherically symmetric BH solutions have been constructed. While the diverse properties, such as thermodynamics, lensing, and shadows of these regular EYM BHs have been explored~\cite{Jawad:2017mwt,Jawad:2018cdh,Liu:2019pov,Rayimbaev:2021vsq,Ali:2025rjs,Jusufi:2020odz,Kala:2022uog,Zhang:2023oui,Chen:2025ujl}, a comprehensive diagnostic using orbital dynamics and observational data remains necessary.

Accretion flows onto such compact objects offer a robust channel to study the strong-gravity regime, as the emitted radiation reflects the underlying spacetime geometry. The soft X-ray continuum from accretion disks~\cite{Bardeen:1972fi} serves as a critical probe for determining the innermost disk radius, which is intrinsically linked to the innermost stable circular orbit (ISCO). Near this boundary, quasi-periodic oscillations (QPOs) observed in X-ray light curves~\cite{lewin2006compact,Motta:2016vwf} provide a direct map of the strong-field dynamics~\cite{syunyaev1972variability}. For low-mass X-ray binaries (LMXBs), QPOs span frequencies from low ($LF < 0.1$ kHz) to high ($HF \sim 0.1-1$ kHz) values~\cite{Stella:1997tc}. Specifically, HFQPOs often appear in paired frequencies with a stable $3:2$ ratio~\cite{Kluzniak:2001ar}, signifying a potential non-linear resonance within the disk. These oscillations can be modeled through the geodesic motion of test particles by analyzing three fundamental epicyclic frequencies, i.e., orbital ($\nu_{\phi}$), radial ($\nu_{r}$), and vertical ($\nu_{\theta}$), which serve as primary diagnostic tools for the inner accretion flow~\cite{Kluzniak:1990}. By comparing theoretical predictions with observed QPOs, one can impose stringent constraints on BH parameters and deviations from GR~\cite{Bambi:2012pa,Bambi:2013fea,Maselli:2014fca,Ghasemi-Nodehi:2020oiz,Chen:2021jgj,Allahyari:2021bsq,Deligianni:2021ecz,Deligianni:2021hwt,Jiang:2021ajk,
Banerjee:2022chn,Liu:2023vfh,Riaz:2023yde,Rayimbaev:2023bjs,Abdulkhamidov:2024lvp,Jumaniyozov:2024eah,Guo:2025zca, Wu:2025ccc, Rehman:2025hfd}. Focusing on the improvements in forthcoming X-ray timing missions, the Insight-HXMT mission (Hard X-ray Modulation Telescope)~\cite{Lu:2019rru} and the next-generation Einstein Probe~\cite{Yuan:2022fpj} are expected to sharpen these tests, enabling more stringent inferences about accretion physics and the nature of central compact objects in strong gravity.

Independent of these orbital oscillations, the rotation of a compact object induces the Lense-Thirring (LT) effect~\cite{Mashhoon:1984fj}, a manifestation of frame-dragging where the spacetime itself is ``dragged" along with the central mass. This relativistic phenomenon, along with geodetic precession~\cite{deSitter:1916zz}, causes the spin axis of a test gyroscope to precess. While verified in the weak-field limit by missions such as Gravity Probe B~\cite{Everitt:2011hp}, LARES~\cite{Capozziello:2014mea}, and LAGEOS~\cite{Ciufolini:2004rq}, the LT effect becomes profoundly dominant in the strong-gravity regime. Recent studies have investigated LT precession around various BHs~\cite{Chakraborty:2013naa, Bini:2016iym, Wu:2023wld,Zhen:2025nah,Wu:2025xtn}, wormholes~\cite{Chakraborty:2012wv}, and neutron stars~\cite{Chakraborty:2014qba}, to characterize how the central object's mass and angular momentum distribution influence the precession rates of orbiting bodies and gyroscopes.

In this work, we investigate the precession of bound orbits, epicyclic oscillations, and the spin precession of test gyroscopes in a rotating regular magnetic BH spacetime coupled to EYM theory. We uniquely combine theoretical derivation with Markov Chain Monte Carlo (MCMC) simulations, using observational QPO data from five X-ray binaries ($GRO~J1655-40, XTE~J1859+226, H1743-322, XTE~J1550-564,$ and $GRS~1915+105$) to constrain the physical parameter space. Moreover, taking a test gyroscope attached to a stationary observer around, we theoretically explore LT, geodetic, and general spin precession frequencies, which are found significantly affected by the BH parameters.

The paper is structured as follows.
In section~\ref{sec:BH and geodesic}, we briefly review the rotating regular magnetic BH, and construct the timelike geodesic motion in the BH background. Then, we study the precession properties of a test particle by analyzing its bound orbit. In section~\ref{sec:QPO constraints}, we firstly study the dynamics of epicyclic motion of timelike particles and explore how the BH parameters influence the epicyclic frequencies. Then, we use the MCMC simulations to determine the best-fit value of the BH parameters with the use of observational data from five QPO events. In section~\ref{sec:Spin Precession}, we consider the geodesic motion of a test gyro attached to a stationary observer in this BH, and theoretically study the corrections to the gyro's LT precession frequency, geodetic precession frequency, and the general spin precession frequency. The final section contributes to our conclusions and discussions.

\section{Timelike bound orbit in the rotating regular magnetic black hole}\label{sec:BH and geodesic}
In this section, we review the fundamental framework of the nonminimally coupled EYM theory and establish the equations of motion for test-particles in the resulting rotating regular magnetic BH spacetime.

\subsection{Review on the rotating regular magnetic black hole} \label{sec:Background}
The action of the nonminimally coupled EYM theory is given by \cite{Balakin:2015gpq, Liu:2019pov}
\begin{equation}
S=\frac{1}{8\pi}\int d^4x \sqrt{-g} \left[R + \frac{1}{2}F^{(a)}_{\mu\nu}F^{\mu\nu(a)}+ \frac{1}{2}\mathfrak{R}^{\alpha\beta\mu\nu}F^{(a)}_{\alpha\beta}F^{(a)}_{\mu\nu}\right],
\label{Eq: action}
\end{equation}
 where $g$ and $R$ represent the determinant of the metric tensor $g_{\mu\nu}$ and Ricci scalar, respectively. Moreover, the Latin indices in the action having range from $0$ to $3$, and the indices ($a$) ranges from $1$ to $3$. The $SU(2)$ YM field strength $F^{(\alpha)}_{\mu\nu}$ is defined via the vector potentials $A^{(\alpha)}_\mu$ as
\begin{equation}
F^{(\alpha)}_{\mu\nu} = \nabla_\mu A^{(\alpha)}_\nu - \nabla_\nu A^{(\alpha)}_\mu + f^{(\alpha)}_{\;\cdot(b)(c)} A^{(b)}_\mu A^{(c)}_\nu, \label{eq:potential}
\end{equation}
where $\nabla_\mu$ is the covariant derivative and $f^{(\alpha)}_{\;\cdot(b)(c)}$ are the real structure constants of the $SU(2)$ group.
The nonminimal susceptibility tensor $\mathfrak{R}^{\alpha\beta\mu\nu}$, which characterizes the coupling between the gauge and gravitational fields, is defined as
\begin{equation}
\mathfrak{R}^{\alpha\beta\mu\nu}=\frac{q_{1}}{2}R\Big(g^{\alpha\mu}g^{\beta\nu}-g^{\alpha\nu}g^{\beta\mu}\Big)+\frac{q_{2}}{2}\Big(R^{\alpha\mu}g^{\beta\nu}-R^{\alpha\nu}g^{\beta\mu}
+R^{\beta\nu}g^{\alpha\mu}-R^{\beta\mu}g^{\alpha\nu}\Big)+q_{3}{R}^{\alpha\beta\mu\nu},\label{Eq:tensor}
\end{equation}
where $R^{\alpha\beta}$ and $R^{\alpha\beta\mu\nu}$ are the Ricci tensor and Riemannian tensor respectively, while $q_{1}, q_{2}, q_{3}$ are the phenomenological parameters.
This action admits an analytical, regular static spherically symmetric BH solution described by the line element~\cite{Balakin:2015gpq}
\begin{equation}
    ds^{2}= -N(r)dt^{2}+ \frac{1}{N(r)}dr^{2}+ r^{2}(d\theta^{2}+\sin^{2}\theta d\phi^{2}),
\end{equation}
with the redshift function
\begin{equation}
    N(r)=1+\Big(\frac{r^4}{r^4+2 \lambda}\Big )\left(-\frac{2 M}{r}+\frac{{Q}^2}{r^2}\right),\label{Eq:static}
\end{equation}
here, $M$ is the ADM mass, $\lambda$ represents the nonminimal coupling parameter, and $Q$ is the magnetic charge of the Wu-Yang gauge field~\cite{Balakin:2006gv}.

\begin{figure}[ht!]
\centering
\includegraphics[width=7cm]{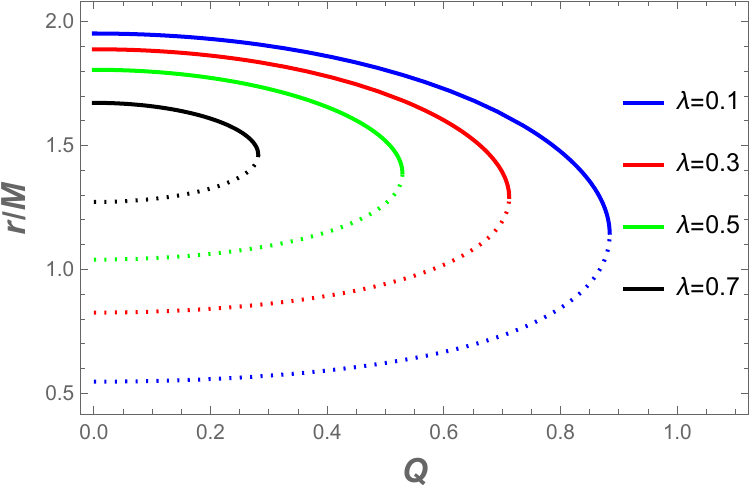}\hspace{1cm}
\includegraphics[width=7cm]{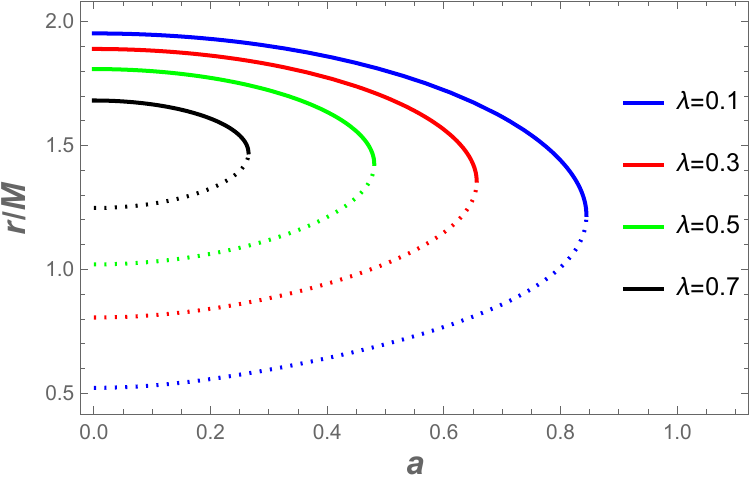}\hspace{1cm}
\caption{The typical plot showing the dependence of the horizon radius on the rotation parameter $a$ and the magnetic charge $Q$, for selected coupling parameters. The solid curves are for the radius of the event horizon ($r_+$) while the dashed curves are for the Cauchy horizon ($r_-$), and their intersection describes the extremal case.}
\label{Fig:horizon}
\end{figure}

Applying the Newman-Janis algorithm, this static solution can be extended to its rotating counterpart in the Boyer-Lindquist coordinates~\cite{Balakin:2015gpq, Liu:2019pov, Jusufi:2020odz}
\begin{equation}
ds^2=-\left(1-\frac{2\Upsilon(r)r}{\Sigma}\right)dt^2 - 2a \sin^2\theta \frac{2\Upsilon(r)r}{\Sigma}dtd\phi + \frac{\Sigma}{\Delta}dr^2 + \Sigma d\theta^2 + \frac{\left((r^2 + a^2)^2 - a^2\Delta \sin^2\theta\right)\sin^2\theta}{\Sigma} d\phi^2, \label{Eq: metric}
\end{equation}
where the associated functions are given by
\begin{equation}\label{Eq:metric-functions}
\Upsilon(r)=\frac{r(1-N(r))}{2},\,\,\,\Delta(r)= r^2 + \left(\frac{r^6}{r^4 + 2\lambda}\right)\left(\frac{-2M}{r}+\frac{Q^2}{r^2}\right) + a^2,\,\,\,\Sigma \equiv r^2 + a^2 \cos^2\theta.
\end{equation}
This metric describes a rotating regular magnetic BH. Note that for $\lambda=0$, the solution reduces to the Kerr-Newman metric with magnetic charge, and for $\lambda=Q=0$, it simplifies to the standard Kerr BH. The horizon radii are determined by the roots of $\Delta(r)=0$. Numerical analysis reveals two positive real roots: the event horizon ($r_+$) and the Cauchy horizon ($r_-$). The dependence of these horizons on the spin $a$, magnetic charge $Q$, and nonminimal parameter $\lambda$ is illustrated in Fig.~\ref{Fig:horizon}. Physically, the nonminimal coupling parameter plays a dominant role, as a larger $\lambda$ systematically suppresses the event horizon $r_+$ and enhances the inner horizon $r_-$ at a fixed $a$ or $Q$, thereby reducing the horizon gap more efficiently.

\subsection{Timelike geodesic and precession of bound orbits}
Analyzing the motion of a test particle in the above rotating regular BH is crucial for understanding the underlying physical processes and their potential observational implications.
The motion of a test-particle of mass $\mu$ is governed by the Lagrangian
\begin{equation}
2\mathcal{S}=g_{\mu \nu} \dot{x}^{\mu}\dot{x}^{\nu} = \delta, \label{Eq:specific-metric}
\end{equation}
where the overdot denotes the derivative with respect to the affine parameter $\tau$, and $\delta = -1$ for timelike geodesics. The stationarity and axisymmetry of the metric yield two constants of motion: the specific energy $E$ and the specific angular momentum $L$,
\begin{eqnarray}
E = -p_t = -(g_{tt}\dot{t} + g_{t\phi}\dot{\phi}), \quad L = p_\phi = g_{t\phi}\dot{t} + g_{\phi\phi}\dot{\phi}, \label{Eq:momenta_EL}
\end{eqnarray}
where $p^\mu=\frac{\mathrm{d}x^\mu}{\mathrm{d}\tau }$ is the four-velocity of the test-particle.
Solving for the four-velocity components $\dot{t}$ and $\dot{\phi}$, we obtain
\begin{equation}
\dot{t}=\frac{E g_{\phi\phi} + L g_{t\phi}}{(g_{t\phi})^{2} - g_{tt}g_{\phi\phi}}, \quad \dot{\phi}=-\frac{E g_{t\phi} + L g_{tt}}{(g_{t\phi})^{2} - g_{tt}g_{\phi\phi}}. \label{Eq:velocities}
\end{equation}
Therefore, the angular velocity of the particle as seen by an observer at infinity, $\Omega_{\phi} = d\phi/dt$, is derived from Eq.~\eqref{Eq:velocities} as
\begin{equation}
\Omega_{\phi} = \frac{-\partial_{r} g_{t\phi} \pm \sqrt{(\partial_{r} g_{t\phi})^{2} - \partial_{r} g_{tt}\partial_{r} g_{\phi\phi}}}{\partial_{r} g_{\phi\phi}}, \label{Eq:angularvelocity-value}
\end{equation}
where the $\pm$ signs correspond to prograde and retrograde orbits, respectively.

For simplicity and clarity, our analysis will focus on particle motion in the equatorial plane ($\theta = \pi/2, \dot{\theta} = 0$), such that
\begin{align}
p^{t} &= \frac{g_{\phi\phi}  \mathit{E} + g_{t\phi}  \mathit{L}}{g_{t\phi}^{2} - g_{tt} g_{\phi\phi}}|_{\theta=\frac{\pi}{2}}=\frac{a^2  \mathit{E} \left(2 \lambda +2 M r^3-Q^2 r^2+r^4\right)+a
\mathit{L} r^2 \left(Q^2-2 M r\right)+r^2  \mathit{E} \left(2 \lambda +r^4\right)}{a^2 \left(2 \lambda +r^4\right)-2 M r^5+Q^2 r^4+r^6+2 \lambda  r^2},\label{Eq:pt}
\end{align}
\begin{align}
p^{\phi}= \frac{-g_{t\phi}  \mathit{E} + g_{tt}  \mathit{L}}{g_{t\phi}^{2} - g_{tt} g_{\phi\phi}}|_{\theta=\frac{\pi}{2}}=\frac{a r^2  \mathit{E} \left(2 M r-Q^2\right)+ \mathit{L} \left(2 \lambda
 -2 M r^3+Q^2 r^2+r^4\right)}{a^2 \left(2 \lambda +r^4\right)-2 M r^5+Q^2 r^4+r^6+2 \lambda  r^2}.\label{Eq:pphi}
\end{align}
Utilizing the above formulas in the normalization condition $p^{\mu}p_{\mu}=-1$ for a timelike geodesic, the radial velocity is determined as
\begin{align}
p^r &= \pm \sqrt{ \frac{-1 - g_{tt}(p^t)^2 - 2 g_{t\phi} p^t p^{\phi} - g_{\phi\phi}(p^{\phi})^2}{g_{rr}} }|_{\theta=\frac{\pi}{2}} \notag \\
&= \pm\Bigg(\frac{1}{r^6 + 2\lambda r^2} \Bigg(a^2\left(2Mr^3 E^2 - Q^2 r^2 E^2 + r^4(E^2 - 1) + 2\lambda(E^2-1)\right)+ 2a L r^2 E (Q^2 - 2Mr) \notag \\
&\quad - L^2 \left(2\lambda - 2Mr^3 + Q^2 r^2 + r^4\right)+ 2Mr^5 - Q^2 r^4+ r^2 (E^2 - 1)(2\lambda + r^4)\Bigg)\Bigg)^{\frac{1}{2}},\label{Eq:pr}
\end{align}
here, $``+"$ and $``-"$ correspond to the particle traveling in the same and opposite directions as the BH rotates, respectively.

\begin{figure}[ht!]
\centering
{\includegraphics[width=7cm]{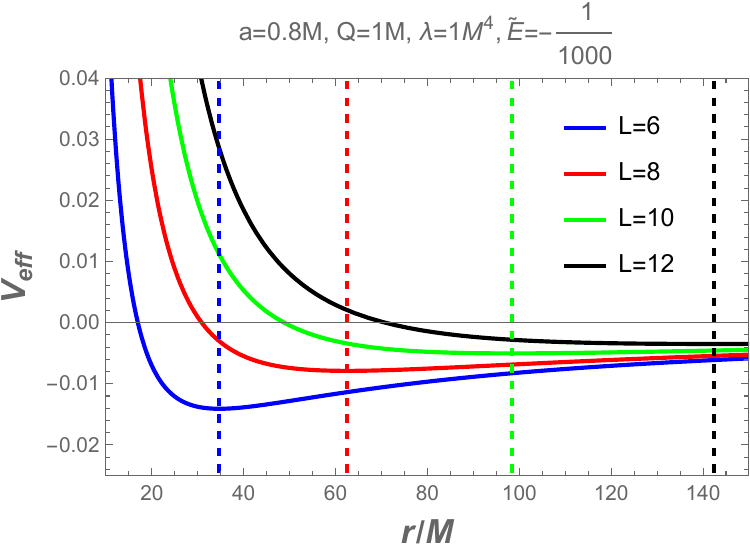}\label{fig:v-r}}\hspace{1cm}
\caption{The effective potential for the regular magnetic BH spacetime as a function of $r/M$ with specific values of $\mathit{\tilde{E}}$, and $\mathit{L}$. The vertical dashed lines indicate the position for each scenario for the minimal effective potential.}\label{fig:vf}
\end{figure}

Subsequently, the effective potential for equatorial timelike geodesics takes the following form,
\begin{align}
V_{\mathrm{eff}}(r)& =\tilde{E} - \frac{1}{2} (p^{r})^2\\
 &=\frac{2 \lambda  \left(L^2-a^2 \left(\mathit{E}^2-1\right)\right)+r^2 \left(r^2 \left(-a^2 \left(\mathit{E}^2-1\right)+\mathit{L}^2+Q^2\right)-2 M r (\mathit{L}
-a \mathit{E})^2+Q^2 (\mathit{L}-a \mathit{E})^2-2 M r^3\right)}{2 r^2 \left(2 \lambda +r^4\right)}, \label{Eq:effective eq}
\end{align}
where $\tilde{E}=\frac{1}{2}\left(\mathit{E}^2-1 \right)$ is the total relativistic energy of the test particle. The physical depiction of the effective potential as a function of the radial coordinate for the timelike geodesics in the regular magnetic BH spacetime, shown in Fig.~\ref{fig:vf}, demonstrates how the effective potential becomes minimal at larger radii for higher values of angular momentum.

%%%%%%%%%%%%%%%%%%%%%%%%%%%%%%%%%
%%%%%%%%%%%%%%%%%%%%%%%%%%%%%%%%%

For the bound orbits, the total energy must not be smaller than the minimum effective potential determined by the system~\cite{Bardeen:1972fi},
\begin{equation}
\left.\frac{d V_{\text{eff}}}{dr}\right|_{r_m} = 0, \quad \left.\frac{d^2 V_{\text{eff}}}{dr^2}\right|_{r_m} > 0.
\end{equation}
Deriving an explicit analytical expression for the minimum radial position $r_{m}$ along with the minimum effective potential $V_{\min}$ proves to be challenging.
Therefore, we display a specimen plot of the effective potential $V_{\mathrm{eff}}$ for a fixed energy value $\tilde{E}=-1/1000$ and a selected angular momentum $\mathit{L}$, numerically locating the positions of $r_{m}$ and $V_{\min}$, as illustrated in Fig.~\ref{fig:vf}. It is evident that test particles with larger angular momentum $\mathit{L}$
exhibit both larger $r_{m}$ and higher $V_{\min}$, a behavior analogous to that observed in Kerr spacetime~\cite{Bambhaniya:2020zno}. Under the constraint
$V_{\min}\leq \tilde{E}<0$ for bound orbits, together with the boundary conditions specified in Eq.~\eqref{Eq:effective eq}, we can analyze the orbital trajectories.
These orbits reveal the variation of the inverse radial coordinate $u \equiv 1/r$ within the equatorial plane as a function of the azimuthal angle $\phi$ as
\begin{equation}
\frac{du}{d\phi} = - u^2 \frac{p^{r}}{p^{\phi}} =-\frac{A_u \sqrt{\frac{R_u}{D_u}}}{B_u},
\end{equation}
with
\begin{equation}
A_u=u^2 \left(a^2+Q^2\right)+2 a^2 \lambda  u^6-2 M u+2 \lambda  u^4+1,
\end{equation}
\begin{equation}
B_u=2 M u (a \mathit{E}-\mathit{L})+Q^2 u^2 (\mathit{L}-a \mathit{E})+2 \lambda  \mathit{L} u^4+\mathit{L},
\end{equation}
\begin{equation}
\begin{split}
R_u
&= (\mathit{E}^2-1)+ u\Big(2M + a^2(\mathit{E}^2-1) - \mathit{L}^2 - Q^2 + u\big( 2(\mathit{L}-a\ \mathit{E})^2 M+ u\big( -(\mathit{L}-a\ \mathit{E})^2 Q^2\\&\quad + 2(\mathit{E}^2-1)\lambda+ 2\lambda u^2(a^2(\mathit{E}^2-1)-\mathit{L}^2) \big)\big)\Big),~~~~D_u=2 \lambda  u^4+1,
\end{split}
\end{equation}
and the corresponding second-order equation is
\begin{equation}
\frac{d^{2}u}{d\phi^{2}}=\frac{4 B_u X_u^2 \left(a^2 u^3 D_u+\left(C_u^{-1}\right)\right)+4 A(u) \left(C_u^{-1}\right)(a \mathit{E}-L)+Z_u}{2 A_u B_u^3 D_u},
\end{equation}
with
\begin{eqnarray}
W_u=2 a^2 \lambda  u^6+Q^2 u^2+2 \lambda  u^4,~~~~
C_u^{\pm }=M u^2+\pm W_u,~~~
X_u=\sqrt{\frac{R_u}{D_u}},
\end{eqnarray}
\begin{equation}
\begin{split}
Y_u
&= \left(
  2\lambda\,u^{5}\!\left(a^{2}(\mathit{E}^{2}-1)-\mathit{L}^{2}\right)
  + a^{2}(\mathit{E}^{2}-1)\,u
  + u^{3}\!\left(2(\mathit{E}^{2}-1)\lambda - Q^{2}(\mathit{L}-a\mathit{E})^{2}\right)
  \right.\\
&\quad \left.
  + 2M\,u^{2}(\mathit{L}-a\mathit{E})^{2}
  - \mathit{L}^{2}u
  - Q^{2}u
\right),
\end{split}
\end{equation}
\begin{equation}
Z_u=A_u B_u \left(2 u D_u \left(2 a^2 \lambda  u^4+Q^2+2 \lambda  u^2\right)+u^2 \left(2 (a \mathit{E}-\mathit{L})^2+\left(C_u^{+}\right)+D_u Y_u\right)+u^2 D_u \left(2 M+Y_u\right)\right).
\end{equation}
\begin{figure}[ht!]
\centering
{\includegraphics[width=7cm]{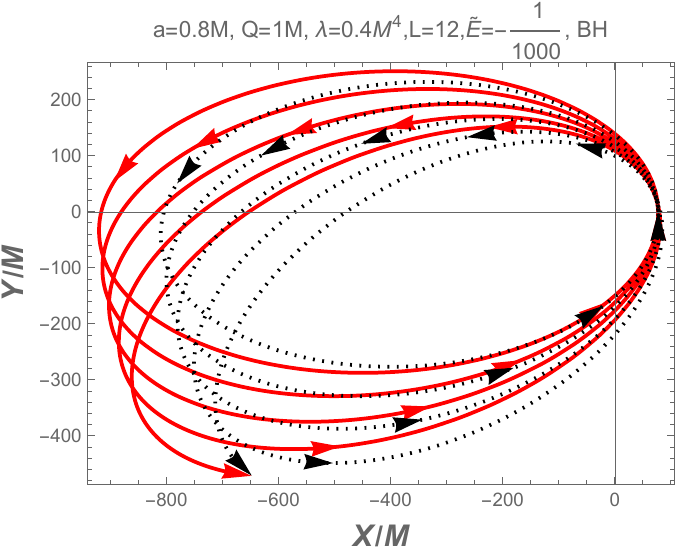}}\hspace{1cm}
\caption{The shape of the precession orbit for a test-particle around a rotating regular magnetic BH spacetime. The dotted black curves indicate the Kerr BH result.}\label{fig:bo}
\end{figure}

By numerically integrating the orbital equation, we reconstruct the geometric configuration of the bound orbit for a test particle in free fall within the regular magnetic BH with spin parameter $a=0.8M$, as shown in Fig.~\ref{fig:bo}. Other parameters used in these calculations are $L=12$, $Q=1M$, $\lambda=0.4M^{4}$, and $\tilde{E}=-1/1000$. In the same figure, we also display the corresponding orbital solutions for the standard Kerr spacetime (black dotted curves) with $\lambda=Q=0$. This comparison highlights the differences between BHs in the regular magnetic spacetime geometry and those in the standard Kerr spacetime~\cite{Bardeen:1972fi}.

%%%%%%%%%%%%%%%%%%%%%%%%
%%%%%%%%%%%%%%%%%%%%%%%%
\section{Parameter constraints through QPO events} \label{sec:QPO constraints}
In this section, we analyze the periastron precession of the circular orbit, characterize both the LT precession and periastron precession in the context of the relativistic precession model (RPM). In this model, the accretion disk is composed of matter moving along nearly circular orbits under the influence of the strong gravitational field of the central compact object, and QPOs observed in X-ray binaries are interpreted as a natural outcome of the motion of matter around the compact object. The characteristic frequencies raised from small perturbations away from circular motion are directly related to the motion of test-particles orbiting in the accretion disk around this BH. Its structure and emission properties are determined by the spacetime geometry, which encodes the gravitational potential and frame-dragging effects specific to the BH metric. Then we further incorporate QPO events observed in X-ray binaries and employ MCMC methods to constrain the parameters of the rotating regular BH.

\subsection{Small perturbation of timelike circular orbit}
In order to model the QPO phenomena of accretion disk by the three fundamental frequencies of the massive particle orbiting the cental object, we introduce small perturbations around the circular orbit with radius $r=r_0$ in equatorial plane, which means the effective potential must satisfy the following conditions:
\begin{equation}\label{eq:Veff}
V_{\mathrm{eff}}(r_0)=0,  \quad \partial_r V_{\mathrm{eff}}(r_0)=0.
\end{equation}
The perturbed coordinates are expressed as
\begin{equation}\label{eq:per}
r(t)=r_0+\delta r(t), \quad \theta(t)=\frac{\pi}{2}+\delta \theta(t),
\end{equation}
where $\delta r(t)$ and $\delta \theta(t)$ are small perturbations from the circular orbit. These perturbations are governed by the following equations of motion:
\begin{equation}\label{eq:perEq}
\frac{d^2 \delta r(t)}{d t^2}+\Omega_r^2 \delta r(t)=0, \quad \frac{d^2 \delta \theta(t)}{d t^2}+\Omega_\theta^2 \delta \theta(t)=0,
\end{equation}
where $\Omega_r$ and $\Omega_\theta$ can be expressed as \cite{Ryan:1995wh,Doneva:2014uma}
\begin{equation}
\Omega _r= \Big(\frac{1}{2 g_{rr}}\Big(\tilde{X}^2\partial^{2}_{r}(\frac{g_{\phi\phi}}{g_{tt}g_{\phi \phi}-g_{t\phi}^{2}})-2\tilde{X}\tilde{Y}\partial^{2}_{r} (\frac{g_{t\phi}}
{g_{tt}g_{\phi \phi}-g_{t\phi}^{2}}) + \tilde{Y}^{2}\partial^{2}_{r} (\frac{g_{tt}}{g_{tt}g_{\phi \phi}-g_{t\phi}^{2}})\Big)\Big)^{\frac{1}{2}},\label{Eq:radial}
\end{equation}
\begin{equation}
\Omega _\theta= \Big(\frac{1}{2 g_{\theta\theta}} \Big(\tilde{X}^2\partial^{2}_{\theta}(\frac{g_{\phi\phi}}{g_{tt}g_{\phi \phi}-g_{t\phi}^{2}})-2\tilde{X}\tilde{Y}\partial^{2}_{\theta}
 (\frac{g_{t\phi}}{g_{tt}g_{\phi \phi}-g_{t\phi}^{2}}) + \tilde{Y}^{2}\partial^{2}_{\theta} (\frac{g_{tt}}{g_{tt}g_{\phi \phi}-g_{t\phi}^{2}})\Big)\Big)^{\frac{1}{2}},\label{Eq:vertical}
\end{equation}
with the quantities $\tilde{X}$, and $\tilde{Y}$ defined by
\begin{equation}
\tilde{X}= g_{tt}+ g_{t\phi}\Omega _\phi,\,\,\,\,\,\ \text{and}\,\,\,\,\,\ \tilde{Y}=g_{t\phi}+ g_{\phi\phi}\Omega _\phi. \label{Eq:XY}
\end{equation}

Finally, the radial epicyclic frequency $\nu_{r}$, the vertical epicyclic frequency $\nu_{\theta}$, and the orbital epicyclic frequency $\nu_{\phi}$ are given in standard physical units as
\begin{equation}
\nu_{\iota}= \frac{c^{3}}{2 \pi G M}\Omega _\iota,~~~~~ \text{where}~~ \iota=(r, \theta, \phi).  \label{Eq:Hz}
\end{equation}
%%%%%%%%%%%%%%%%%%%%%%%%%%%%%%%%%%
%%%%%%%%%%%%%%%%%%%%%%%%%%%%%%%%%%
%%%%%%%%%%%%%%%%%%%%%%%%%%%%%%%%%%
\begin{figure}[ht!]
\centering
\includegraphics[width=5.5cm]{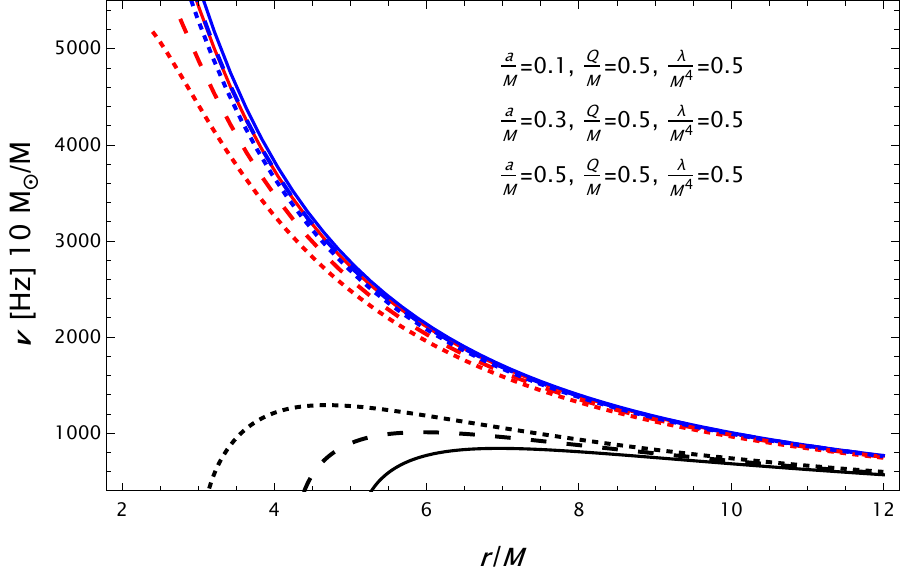}\hspace{1cm}
\includegraphics[width=5.5cm]{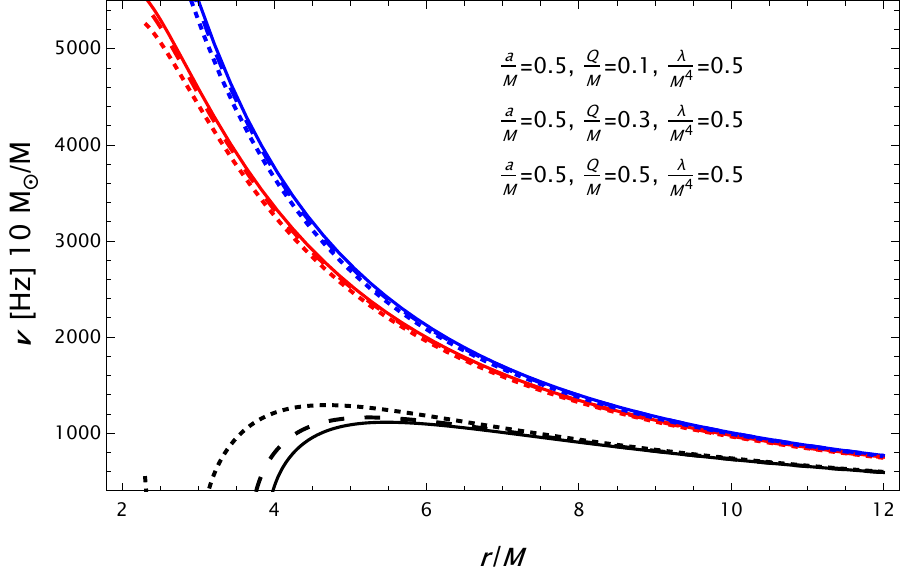}
\includegraphics[width=5.5cm]{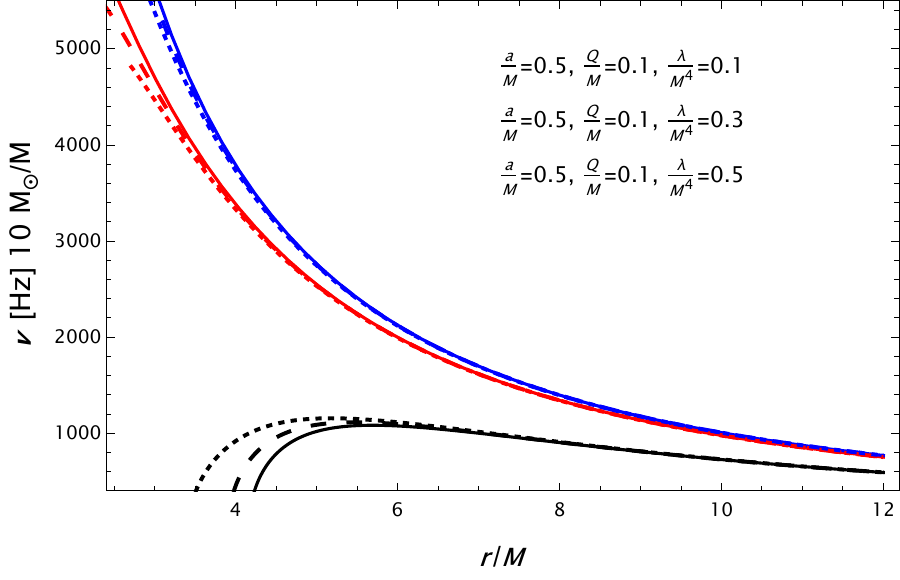}
\caption{The trends of the radial (black curves), vertical (red curves), and orbital epicyclic (blue curves) frequencies as functions of the ratio $r/M$ are analyzed for various values of the spin, magnetic charge, and the nonminimal coupling parameters, respectively.}\label{fig:vf-overall}
\end{figure}
In Fig.~\ref{fig:vf-overall}, we analyze the relationships among the radial epicyclic frequency $\nu_{r}$ (black curves), the vertical epicyclic frequency $\nu_{\theta}$ (red curves), and the orbital frequency $\nu_{\phi}$ (blue curves), together with their corresponding ratios, as functions of the dimensionless radial coordinate $r/M$ for different values of $a/M$, $Q/M$, and $\lambda/M^{4}$.

Initially, the left panel illustrates the convergence of the vertical and orbital frequencies, both of which decrease monotonically as $a/M$ increases from $0.1$ (solid curves) to $0.5$ (dotted curves), and also decrease as the range of $r/M$ increases. In particular, with increasing $a/M$, the vertical frequency (red curves) exhibits a more pronounced reduction than the orbital frequency. Moreover, as $a/M$ increases, the radial epicyclic frequency (black curves) is shifted toward smaller values of $r/M$, exhibiting an opposite trend characterized by a decrease. This behavior highlights the nontrivial influence of the BH spin on the structure of epicyclic modes and has important implications for the stability and oscillatory motion of matter in the strong-gravity regime.
The central panel depicts the variation of all three characteristic frequencies with respect to the dimensionless charge parameter, while $a/M$ and $\lambda/M^{4}$ are held fixed. The resulting behavior remains in agreement with the trends discussed previously.
In the rightmost panel, within a narrow interval of the radial coordinate, both the vertical and orbital frequencies decrease monotonically along the radial direction as $\lambda/M^{4}$ increases. The radial frequency also increases with increasing $\lambda/M^{4}$, although its overall behavior exhibits a decreasing trend along the radial coordinate.
%%%%%%%%%%%%%%%%%%%%%%%%%%%%%%%%%%
%%%%%%%%%%%%%%%%%%%%%%%%%%%%%%%%%%

The triplet frequencies can be more directly compared to observational frequencies measured in standard physical units of Hz by using Eq.~\eqref{Eq:Hz} as follows
\begin{equation}
\nu_{r}=\frac{\Omega_{r}}{2\pi},~~~~~~~\nu_{\theta}=\frac{\Omega_{\theta}}{2 \pi},~~~~~~~ \nu_{\phi}=\frac{\Omega_{\phi}}{2 \pi},   \label{Eq:conversion}
\end{equation}
where
\begin{equation}
\nu_{\phi}=\frac{1}{2\pi}\Big( \frac{\sqrt{M r \left(r^4-6 \lambda \right)-Q^2 \left(r^4-2 \lambda \right)}}{2 \lambda +\text{Ma}_* \sqrt{M r \left(r^4-6 \lambda \right)-Q^2 \left(r^4-2 \lambda \right)}+r^4}\Big),  \label{Eq:keplerian frequency}
\end{equation}

\begin{align}\nonumber
\nu_{r} &=\nu _{\phi }\Big( Q^2 \left(9 M r^8+2 \lambda  r^4 (M-12 r)+16 \lambda ^2 r\right)+M r^2 \left(-60 \lambda ^2-6 M \left(r^7+2 \lambda  r^3\right)+r^8+36 \lambda  r^4\right)-4 Q^4 r^7\notag \\\nonumber
&\quad +8 \text{Ma}_* r \left(M r \left(r^4-6 \lambda \right)-Q^2 \left(r^4-2 \lambda \right)\right)^{3/2}+\text{Ma}_*^2 \left(M \left(-12 \lambda ^2-3 r^8+52 \lambda  r^4\right)+4 Q^2 \left(r^7-6 \lambda  r^3\right)\right)\\
&\quad r \left(2 \lambda +r^4\right) \left(M r \left(r^4-6 \lambda \right)-Q^2 \left(r^4-2 \lambda \right)\right)^{-1}\Big), \label{Eq:radial physical frequency1}
\end{align}
\begin{eqnarray}
\nu_{\theta}=\nu _{\phi }\Big(\frac{2 \text{Ma}_* \left(Q^2-2 M r\right)}{\sqrt{M r \left(r^4-6 \lambda \right)-Q^2 \left(r^4-2 \lambda \right)}}+\frac{\text{Ma}_*^2 \left(-2 \lambda  M+3 M r^4-2 Q^2 r^3\right)}{r \left(M r \left(r^4-6 \lambda \right)-Q^2 \left(r^4-2 \lambda \right)\right)}+1 \Big), \label{Eq:radial physical frequency2}
\end{eqnarray}
here ${a}_*\equiv \frac{a}{M}$. Within the equatorial plane of test particle motion, the radial epicyclic frequency describes oscillations in the radial direction about the mean circular orbit, whereas the vertical epicyclic frequency characterizes oscillations perpendicular to the equatorial plane.

Using the three fundamental frequencies discussed above, we formulate the relativistic periastron precession model for X-ray BH binaries. In this framework, the periastron precession and nodal precession frequencies are denoted by $\nu_{per}$ and $\nu_{nod}$, respectively, and are defined as~\cite{Stella:1999sj}
\begin{equation}
\nu_{per}= \nu _{\phi }-\nu _{r},~~~~~~~~ \text{and}~~~~~~~~ \nu_{nod}= \nu _{\phi }-\nu_{\theta},  \label{Eq:per-nod}
\end{equation}
where $\nu_{per}$ corresponds to the periastron precession of an eccentric orbit, while $\nu_{nod}$ (LT) represents the precession of the orbital plane induced by frame-dragging effects.

\begin{figure}[ht!]
\centering
\includegraphics[width=7cm]{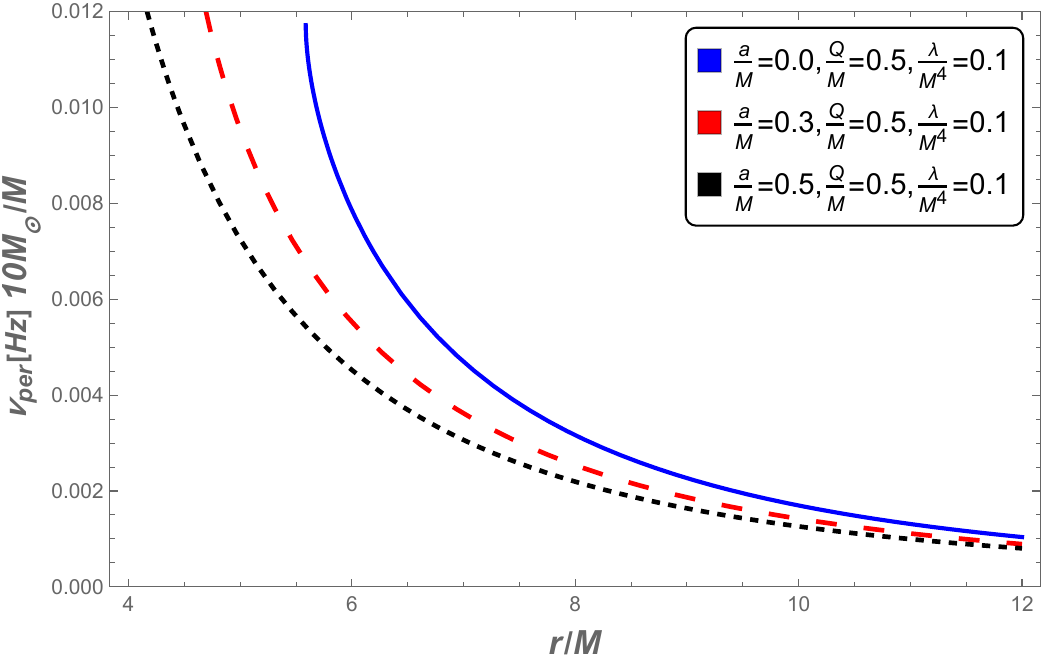}\hspace{1cm}
\includegraphics[width=7cm]{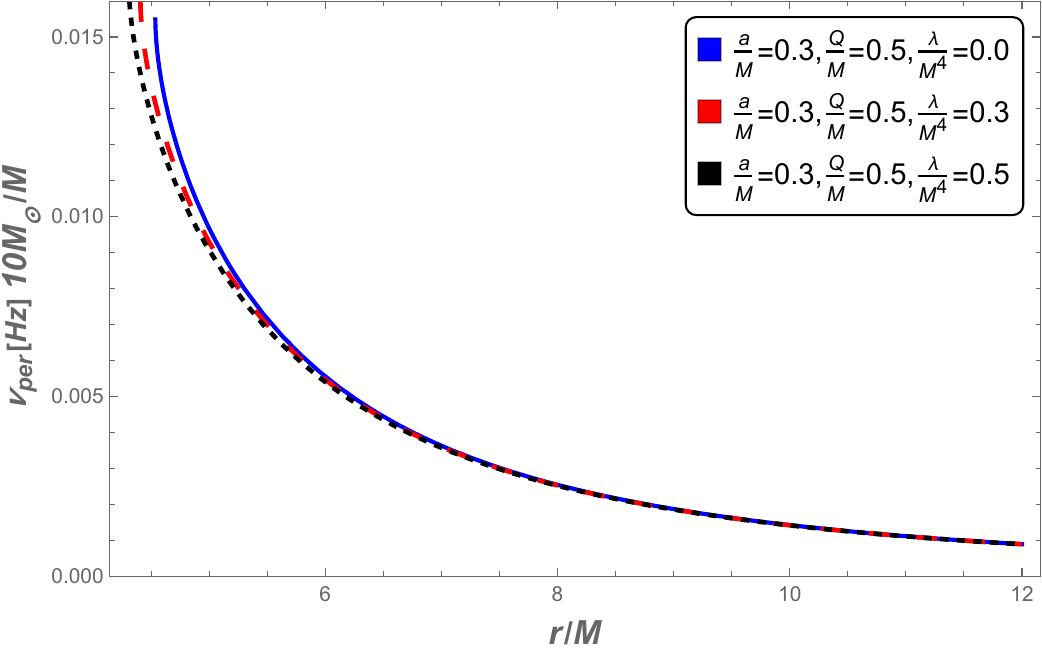}
\includegraphics[width=7cm]{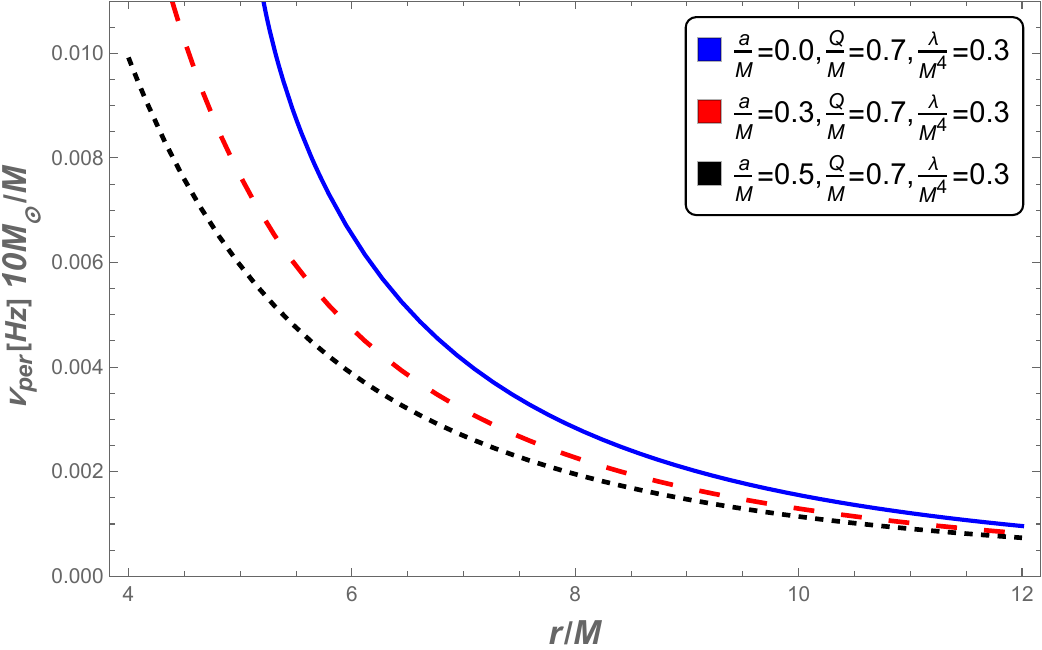}\hspace{1cm}
\includegraphics[width=7cm]{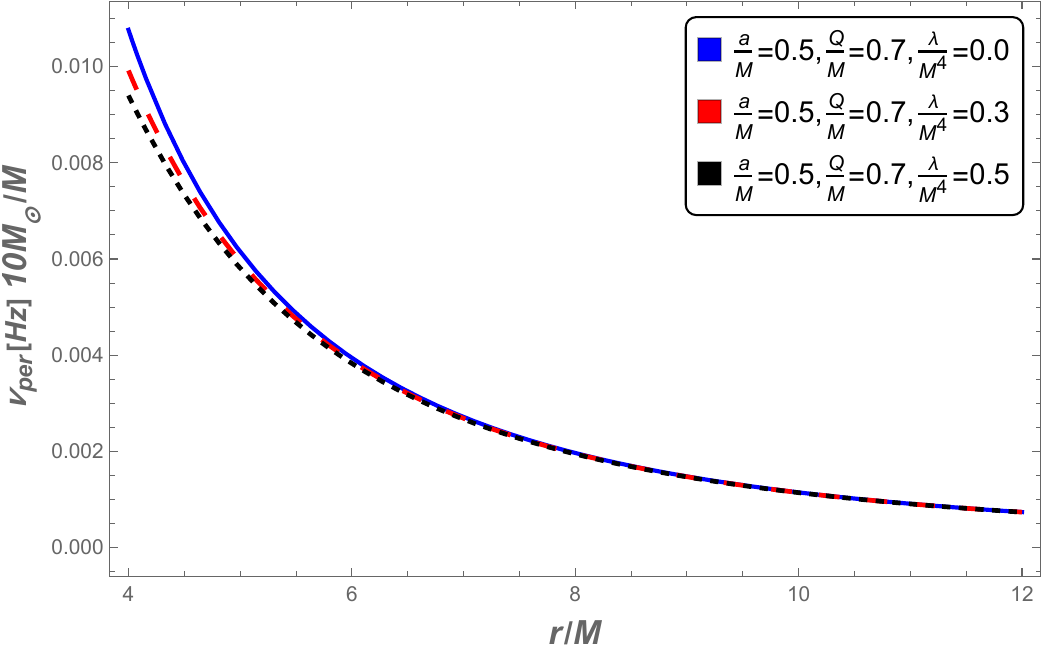}
\caption{The behavior of the periastron precession frequency $\nu_{per}$ is analyzed as a function of the ratio $r/M$ for fixed values of the magnetic charge $(Q = 0.5, 0.7)$. The spin parameter and the nonminimal coupling parameter are varied independently.}\label{fig:per-overall}
\end{figure}
\begin{figure}[ht!]
\centering
\includegraphics[width=7cm]{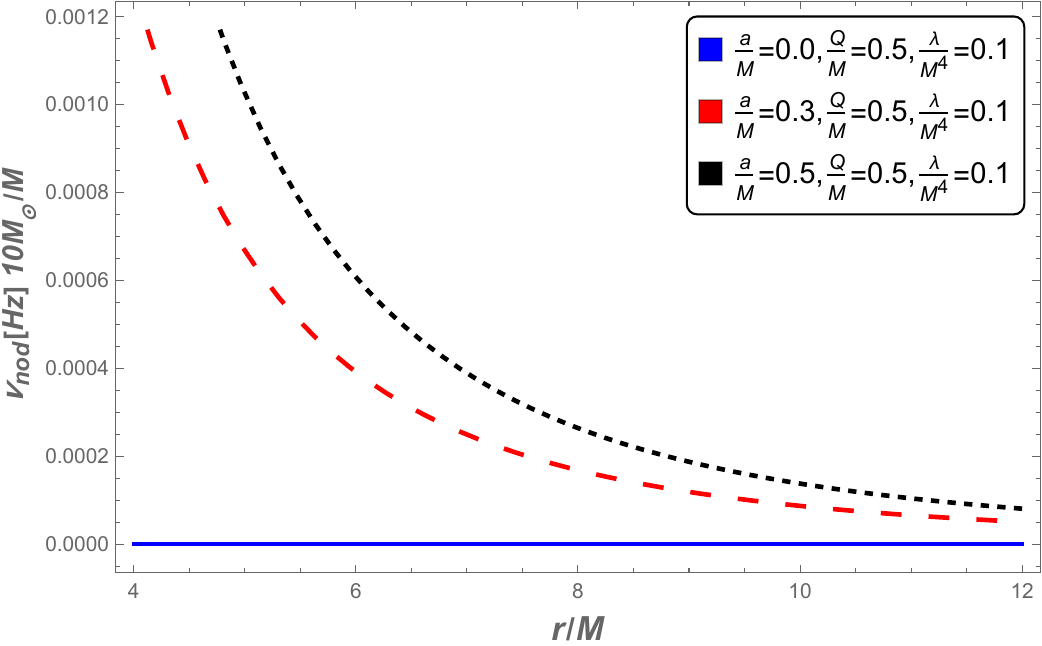}\hspace{1cm}
\includegraphics[width=7cm]{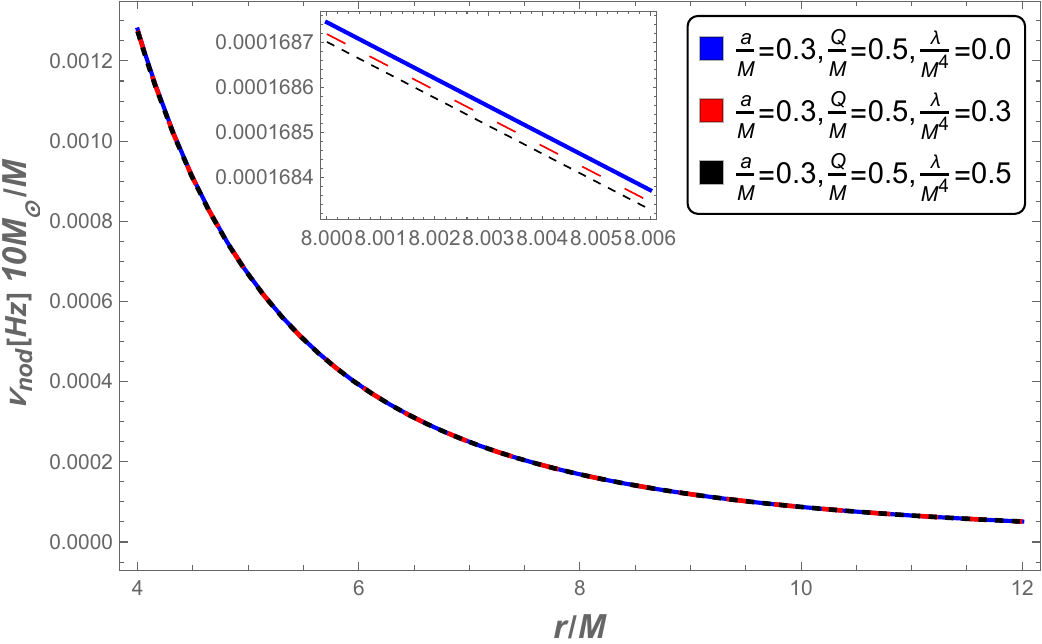}
\includegraphics[width=7cm]{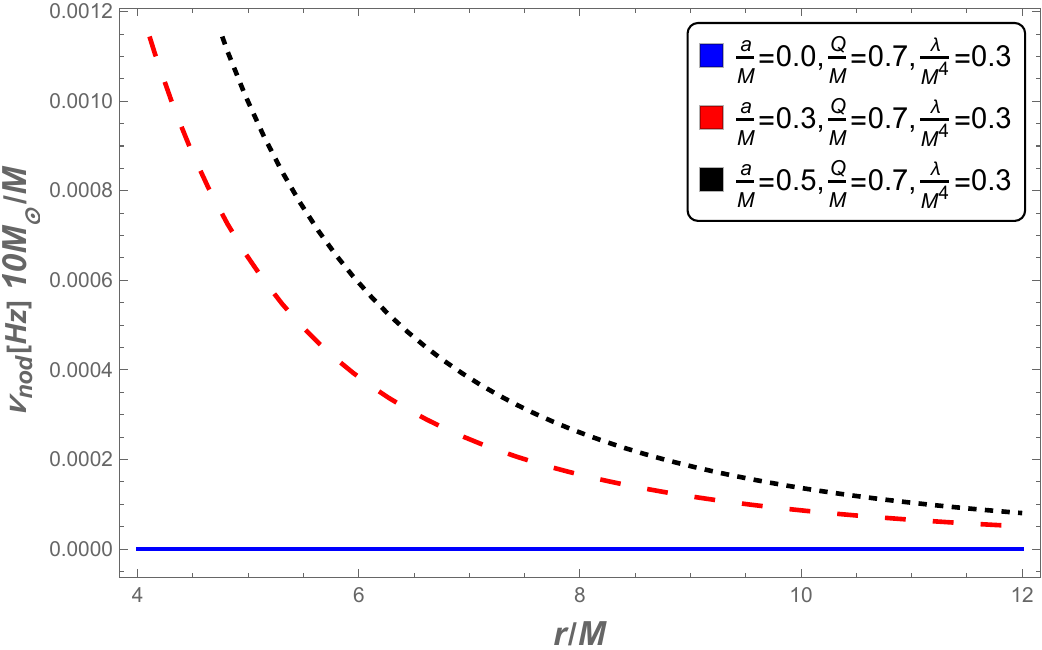}\hspace{1cm}
\includegraphics[width=7cm]{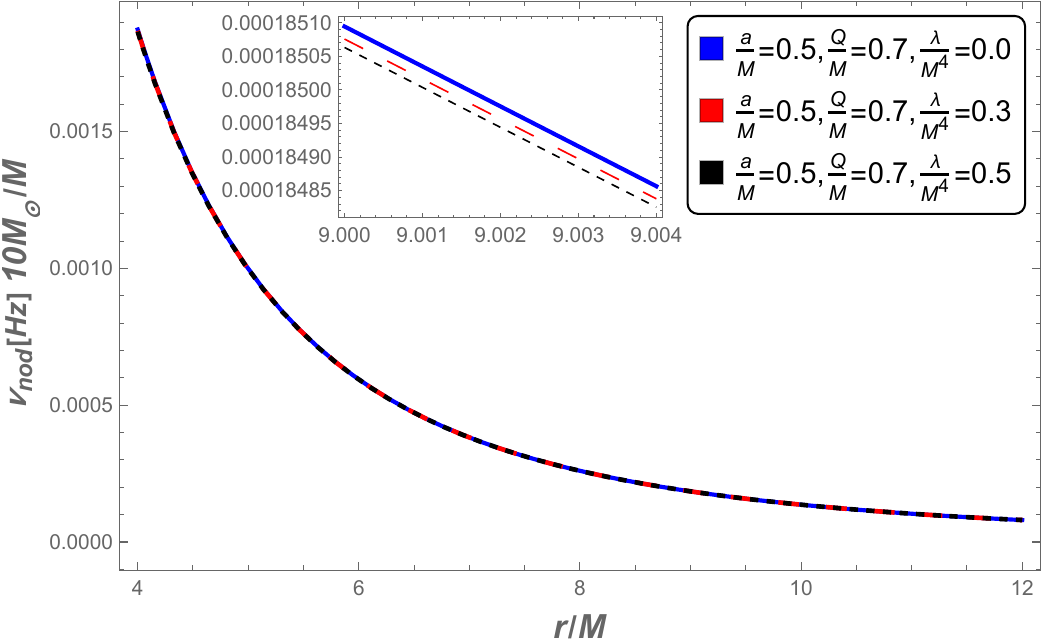}
\caption{The behavior of the nodal precession frequency $\nu_{nod}$ is examined as a function of the ratio $r/M$ for fixed values of the magnetic charge $(Q = 0.5, 0.7)$, while varying the spin parameter and the nonminimal parameter.}\label{fig:nod-overall}
\end{figure}
The radial dependence of the periastron precession frequency $\nu_{per}$ and the nodal precession frequency $\nu_{nod}$ is displayed in Figs.~\ref{fig:per-overall} and~\ref{fig:nod-overall}. These figures are presented to illustrate the influence of the spin parameter as well as the effect of the nonminimal coupling parameter $\lambda/M^{4}$ in the regular magnetic BH spacetime. From the Figs., it is evident that both the periastron precession frequency and the nodal precession frequency exhibit a pronounced monotonic decrease with increasing radial distance, eventually approaching zero asymptotically at spatial infinity.

Specifically, in the left panels of Figs.~\ref{fig:per-overall} and~\ref{fig:nod-overall}, increasing the BH spin parameter leads to a radial suppression of the periastron precession frequency $\nu_{per}$, while simultaneously enhancing the nodal precession frequency $\nu_{nod}$, for fixed values of $\lambda/M^{4}$. In contrast, the right panels of both Figs. indicate that an increase in the nonminimal coupling parameter $\lambda/M^{4}$ leads to a suppression of both the periastron and LT precession frequencies. This suppressive effect is more pronounced for the periastron precession frequency compared to the nodal precession frequency. Furthermore, in the absence of the spin parameter $(a/M = 0)$, the nodal precession frequency vanishes, indicating that no nodal precession frequency occurs in a non-rotating spacetime.

Next, we shall combine the theoretical results and observational data of QPOs, and employ MCMC simulations to explore the space of physical parameters and to constrain the range of the BH parameters.

\subsection{Constraining parameter  through MCMC analysis}
In this subsection, we conduct a comprehensive analysis of QPOs observed in well-studied X-ray binary systems to constrain the parameters of rotating regular magnetic BH spacetime. Our study focuses on five prominent QPO events observed from exclusively analyzed X-ray binary sources, namely $GRO~J1655-40$, $XTE~J1859+226$, $H1743-322$, $XTE~J1550-564$, and $GRS~1915+105$, as summarized in Table~\ref{table:Xray-sor}. By combining the theoretical predictions based on geodesic motion with observational QPO data, we employ MCMC simulations to systematically explore the multidimensional parameter space and accurately determine the constraints on the dimensionless parameters of this BH spacetime. The corresponding best-fit values and physically viable parameter ranges are subsequently reported. The MCMC analysis is performed using the \text{emcee} sampler, following the methodology outlined in Ref.~\cite{ForemanMackey2013}.

The posterior probability is described according to Bayes' theorem as
\begin{equation}
\mathcal{P}(\Theta | \mathcal{D}, \mathcal{M}) = \frac{P(\mathcal{D} | \Theta, \mathcal{M}) \, \pi(\Theta | \mathcal{M})}{P(\mathcal{D} | \mathcal{M})},
\end{equation}
where $\pi(\Theta | \mathcal{M})$, $P(\mathcal{D} | \Theta, \mathcal{M})$ and $P(\mathcal{D} | \mathcal{M})$ present the prior distribution on the parameters, the likelihood, and the normalization factor, respectively.

The truncated Gaussian prior distribution is adopted for the subsequent analysis and is subject to the constraints
\begin{equation}
\pi(\theta_i) \sim \exp\!\left[-\tfrac{1}{2}\left(\frac{\theta_i-\theta_{0,i}}{\sigma_i}\right)^2\right],
\qquad \theta_{\mathrm{low},i} < \theta_i < \theta_{\mathrm{high},i},
\end{equation}
here, the prior corresponds to a Gaussian distribution for the parameters $\theta_i=\big[M,\, a/M,\, r/M\big]$, with $\sigma_i$ denoting the standard deviations that quantify the uncertainty associated with each parameter. For the analysis of the additional parameters $\big[\,|Q|/M,\, \lambda/M^{4}\,\big]$, a uniform prior distribution is considered.

Three distinct data sets are employed in our MCMC analysis, based on the orbital, periastron, and nodal precession frequencies derived in the preceding section.
As a result, the likelihood function $\mathcal{L}$ consists of three parts and can be expressed as
\begin{equation}
\log \mathcal{L} = \log \mathcal{L}_{\text{obt}} + \log \mathcal{L}_{p} + \log \mathcal{L}_{n},
\end{equation}
 where $\mathcal{L}_{\text{obt}}$, $\mathcal{L}_{p}$, and $\mathcal{L}_{n}$ denote the likelihoods associated with the orbital, periastron precession, and nodal precession frequencies, respectively.

Explicitly,
\begin{equation}
\log \mathcal{L} = -\frac{1}{2} \sum_i \frac{\left( \nu_{\phi,\text{obs}}^i - \nu_{\phi,\text{th}}^i \right)^2}{\left( \sigma_{\phi,\text{obs}}^i \right)^2}-\frac{1}{2} \sum_i \frac{\left( \nu_{\text{p,obs}}^i - \nu_{\text{p,th}}^i \right)^2}{\left( \sigma_{\text{p,obs}}^i \right)^2}-\frac{1}{2} \sum_i \frac{\left( \nu_{\text{n,obs}}^i - \nu_{\text{n,th}}^i \right)^2}{\left( \sigma_{\text{n,obs}}^i \right)^2},
\end{equation}
where $\nu_{\phi,\text{obs}}^i$, $\nu_{\text{p,obs}}^i$, and $\nu_{\text{n,obs}}^i$ denote the observed orbital, periastron precession, and nodal precession frequencies, while $\nu_{\phi,\text{th}}^i$, $\nu_{\text{p,th}}^i$, and $\nu_{\text{n,th}}^i$ represent the corresponding theoretical predictions. Moreover, $\sigma_{\text{x,obs}}^i$ denotes the statistical uncertainty associated with the respective observed quantities.

Following this setup, and using the prior distribution in Table~\ref{table:Xray-para}, we generate $10^{5}$ random samples for each parameter to fully explore the physically admissible parameter space and to determine the posterior distribution with specified boundaries for the BH parameters $|Q|/M$ and $\lambda/M^{4}$. We choose the same Gaussian priors for the parameters $(M, a/M, r/M)$ as those used in \cite{Rehman:2025hfd,Wu:2025xtn,Motta:2013wga,Motta:2022rku,Ingram:2014ara,Remillard:2002cy,Remillard:2006fc}. While for the BH parameters, we adopt uniform priors in the regimes $|Q|/M \in [0,0.4]$ and $\lambda/M^{4} \in [0,1]$, thereby confining the analysis to a parameter region associated with moderate deviations from the standard Kerr spacetime geometry. Within this restricted domain, the RPM remains mathematically well defined, real, and monotonic, which permits the existence of simultaneous and physically consistent solutions for all three QPO frequencies.\\
\begin{table}[ht!]
\caption{The analysis of selected X-ray binaries for QPOs aims to discuss their mass, orbital, periastron precession, and nodal precession frequencies.} % title of Table
\centering % used for centering table
\begin{tabular}{c c c c c c c c c c } % centered columns (7 columns)
\hline %inserts double horizontal lines
\hline %inserts double horizontal lines
\\
Source&~~~GRO J1655-40 \cite{Motta:2013wga} &~~~XTE J1859+226 \cite{Motta:2022rku} & ~~~H1743-322 \cite{Ingram:2014ara}&~~~XTE J1550-564 \cite{Remillard:2002cy}&~~~GRS 1915+105 \cite{Remillard:2006fc} \\ [0.5ex]
\hline %inserts double horizontal lines
\hline %inserts double horizontal lines
\\
$M~(M_{\bigodot})$&     ~~~~~~~ $5.4\pm0.3$&~~~~~~~$7.85\pm0.46$&~~~~~~~$\gtrsim 9.29$&~~~~~~~$9.1\pm0.61$&~~~~~~~$12.4^{+2.0}_{-1.8}$ \\
\\

$\nu_{\phi}$~(Hz)&     ~~~~~~~ $441\pm2$&~~~~~~~$227.5^{+2.1}_{-2.4} $&~~~~~~~$240\pm3  $&~~~~~~~$276\pm3$&~~~~~~~$168\pm3$ \\
\\

$\nu_{per}$~(Hz)&      ~~~~~~~ $298\pm4$&~~~~~~~$128.6^{+1.6}_{-1.8} $&~~~~~~~$165^{+9}_{-5}$&~~~~~~~$184\pm5$&~~~~~~~$113\pm5$ \\
\\

$\nu_{nod}$~(Hz)&      ~~~~~~~ $17.3\pm0.1$&~~~~~~~$3.65\pm0.01$&~~~~~~~$9.44\pm0.02 $&~~~~~~~$-$&~~~~~~~$-$ \\

\hline %inserts double horizontal lines
\hline %inserts double horizontal lines
\label{table:Xray-sor}
\end{tabular}
\label{table:nonlin}
\end{table}
%%%%%%%%%%%
%%%%%%%%%%%
\begin{table}[ht!]
\caption{ The Gaussian prior is chosen for the parameters $(M, a/M, r/M)$, and the uniform priors are chosen for the $(|Q|/M, \lambda/M^{4})$ for the regular magnetic BH parameters inferred from QPOs in the specified X-ray binaries. } % title of Table
\centering % used for centering table
\begin{tabular}{c c c c c c c c c c } % centered columns (7 columns)
\hline %inserts double horizontal lines
\hline %inserts double horizontal lines
\\
Source&~~~GRO J1655-40 &~~~XTE J1859+226 & ~~~H1743-322&~~~XTE J1550-564&~~~GRS 1915+105 \\
\hline %inserts double horizontal lines

Parameters&~~~ $\mu,~\sigma$ &~~~$\mu,~\sigma$ & ~~~$\mu,~\sigma$&~~~$\mu,~\sigma$&~~~$\mu,~\sigma$ \\

\hline %inserts double horizontal lines
\hline %inserts double horizontal lines
\\
$M (M_{\bigodot})$&  ~~~ $5.307,~0.066$&~~~~~$7.85,~0.46$&~~~~~$9.29,~0.46$&~~~~~$9.10,~0.61$&~~~~~$12.41,~0.62$ \\
\\

$a/M$&           ~~~ $0.286,~0.003$&~~~~~   ~$0.149,~0.005$&~~~~~ $0.27,~0.013$&~~~~~ $0.34,~0.007$&~~~~~ $0.29,~0.015$ \\
\\

$r/M$&              ~~~ $5.677,~0.035$&~~~~~$6.85,~0.18$&~~~~~$5.55,~0.27$&~~~~~$5.47,~0.12$&~~~~~$6.10,~0.30$ \\
\\

$ |Q|/M$&              ~~~~~ Uniform$[0, 0.4]$&~~~~~Uniform$[0, 0.4]$&~~~~~Uniform$[0, 0.4]$&~~~~~Uniform$[0, 0.4]$&~~~~~Uniform$[0, 0.4]$ \\
\\

$ \lambda/M^{4}$&     ~~~~~ Uniform$[0, 1]$&~~~~~Uniform$[0, 1]$&~~~~~Uniform$[0, 1]$&~~~~~Uniform$[0, 1]$&~~~~~Uniform$[0,1]$ \\
\hline %inserts double horizontal lines
\hline %inserts double horizontal lines
\label{table:Xray-para}
\end{tabular}
\end{table}
%%%%%%%%%%%
%%%%%%%%%%%

\begin{figure}[ht]
\centering
\includegraphics[width=9cm]{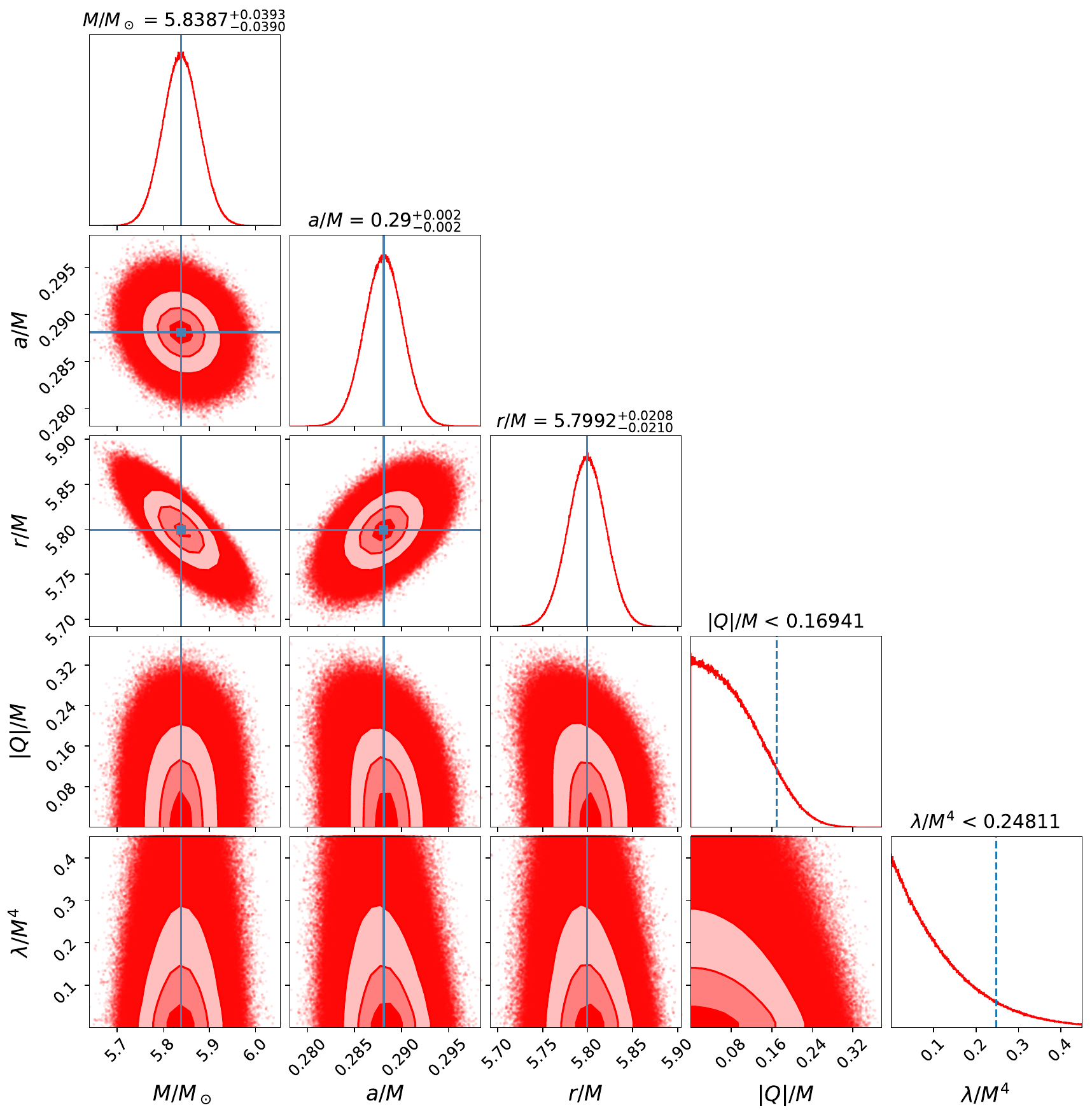}
\caption{Marginalized distribution constraints on the model parameters of the rotating regular BH associated with the $GRO~J1655-40$, inferred current observations of QPOs within the RPM.}\label{Fig:MC1}
\end{figure}

\begin{figure}[ht]
\centering
{\includegraphics[width=7cm]{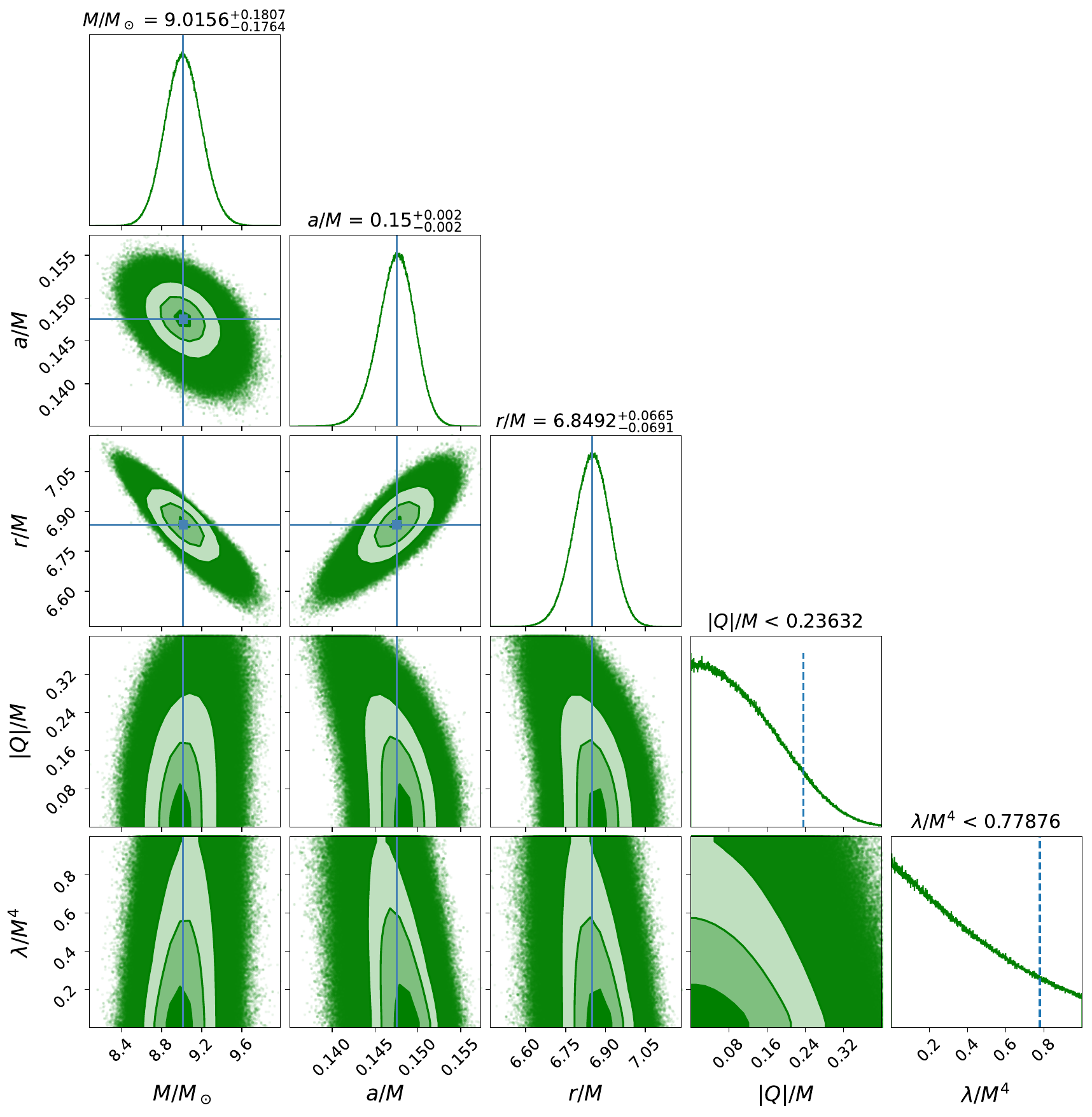}}\hspace{0.3cm}
{\includegraphics[width=7cm]{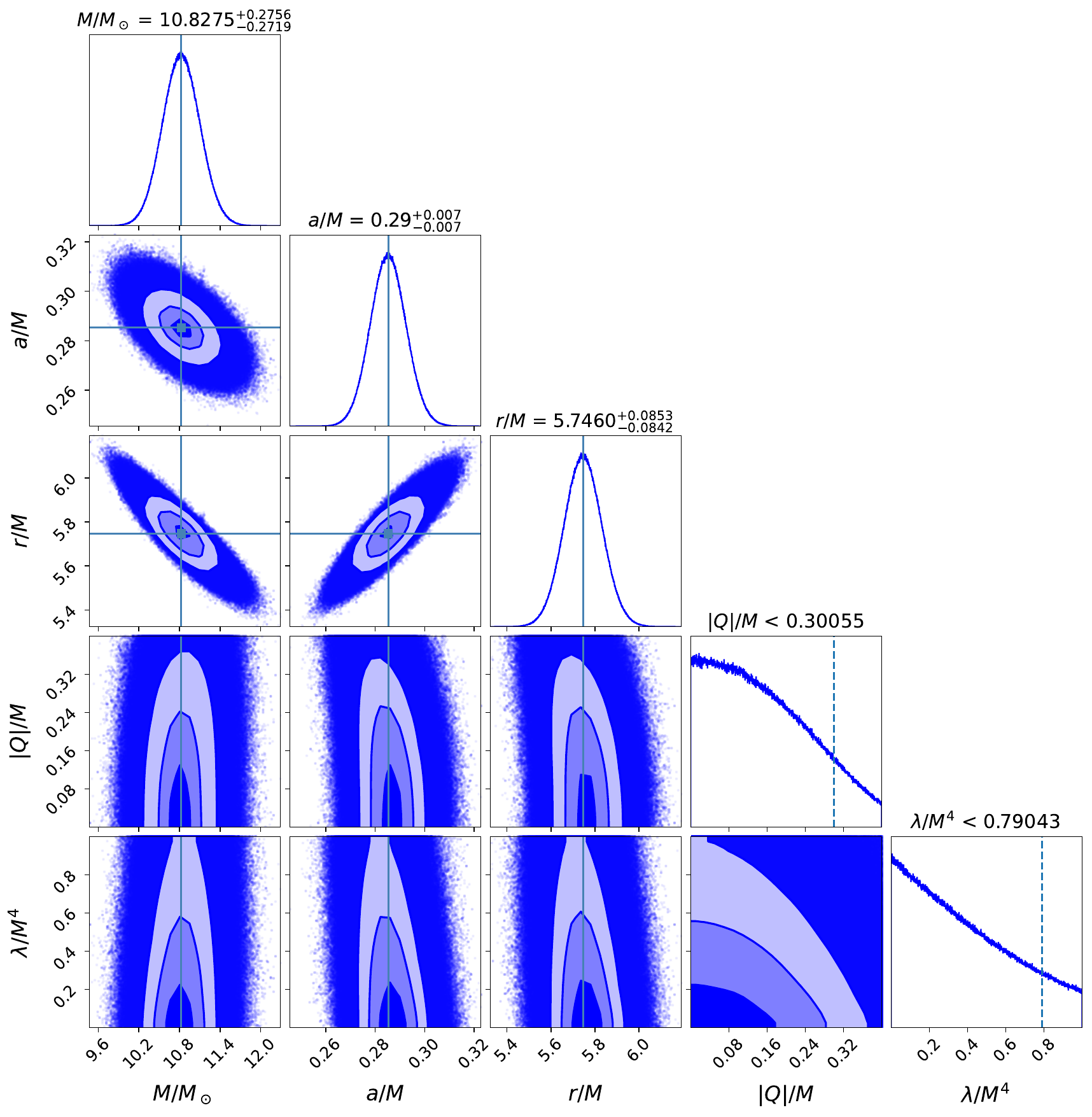}}\\
{\includegraphics[width=7cm]{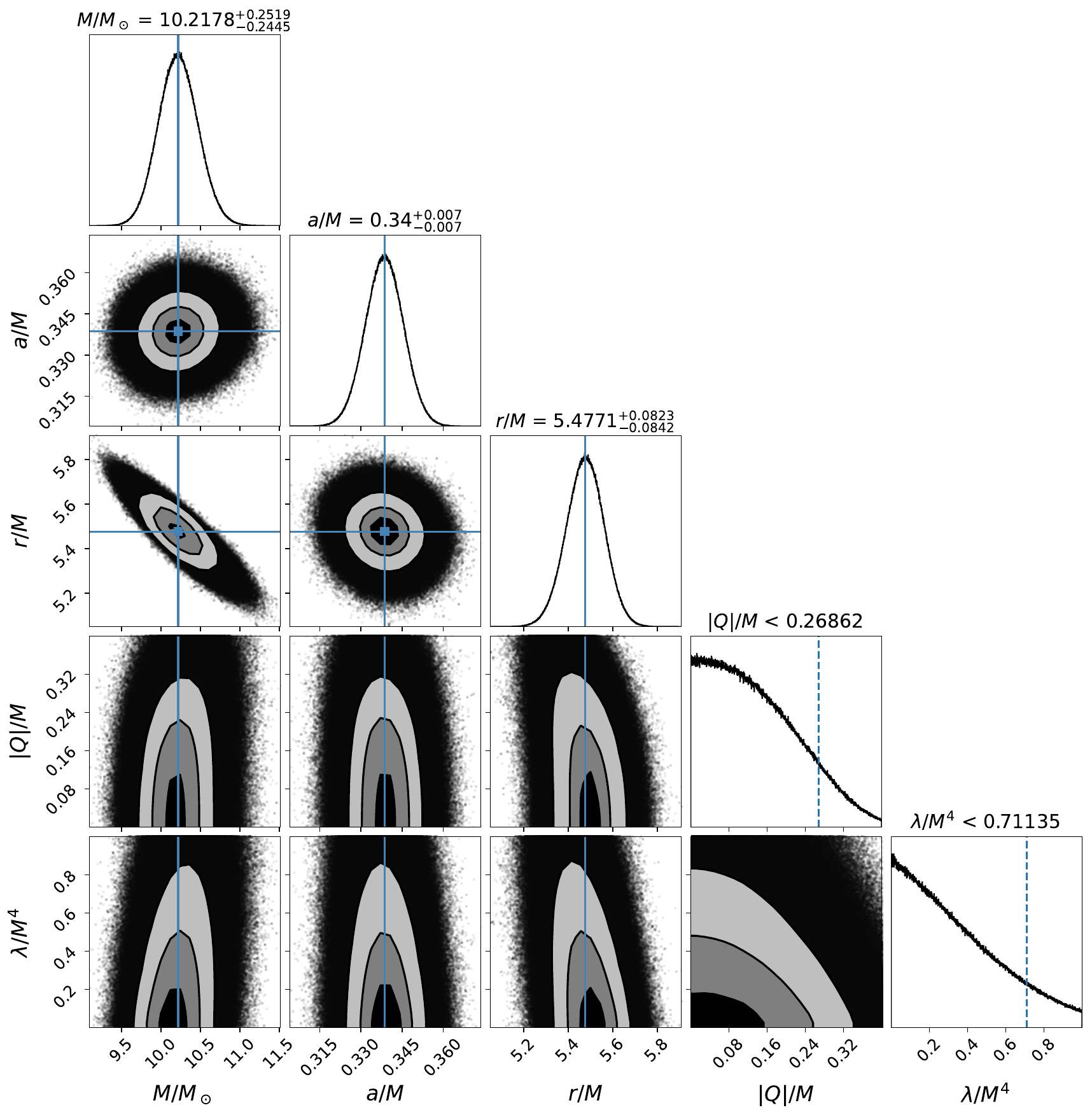}}\hspace{0.3cm}
{\includegraphics[width=7cm]{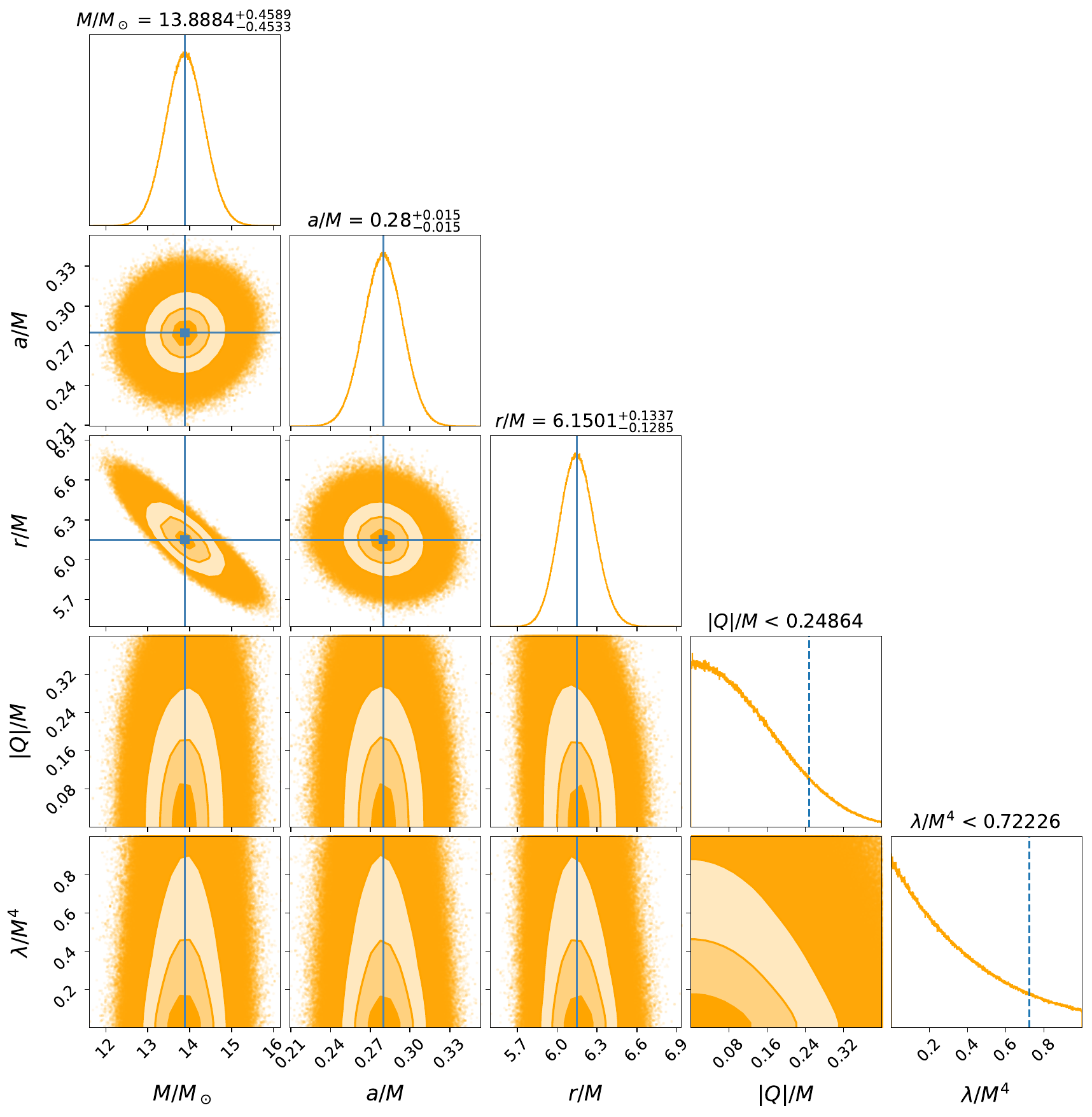}}
\caption{Marginalized distribution constraints on the model parameters of the rotating regular magnetic BH associated with the $XTE~J1859+226$ (green contours), $H1743-322$ (blue contours), $XTE~J1550-564$ (black contours), and $GRS~ 1915+105$ (yellow contours), inferred from current observations of QPOs within the RPMs.}\label{Fig:MC2}
\end{figure}

This analysis rigorously investigates the five-parameter space ($M, a/M, r/M, |Q|/M, \lambda/M^{4}$) describing a regular magnetic BH through MCMC posterior analysis. Figs.~\ref{Fig:MC1} and~\ref{Fig:MC2} display the full posterior distributions of our model parameters within the RP framework. These best outcomes correspond to the applications of the preceding model to the five X-ray binary QPO events, respectively. In the contour plots, the shaded regions denote the $68\%$ and $90\%$ confidence levels (C.L.) of the posterior probability density distributions that clearly highlight the statistically significant parameter constraints. The dashed vertical lines denote the $90\%$ C.L. for the parameters $|Q|/M$ and $\lambda/M^{4}$, respectively.
These figures demonstrate the best constraint results on the model parameters of the rotating regular magnetic BH obtained from current QPO observations within the RPM.
Table~\ref{table:Best-fit} summarizes the best-fit estimates and robust parameter intervals, offering a concise overview of the inferred BH characteristics and constraints.
%%%%%%%%%%%%%%%%%%%%%%%%%%%%%%%%%%%%%%%%%%%%%%%%%%%%%%%%%%%%%%%%%%%%%
\begin{table}[ht!]
\caption{The best-fit values of the rotating regular magnetic BH parameters inferred from QPOs in the specified X-ray binaries.} % title of Table
\centering % used for centering table
\begin{tabular}{c c c c c c c c c c } % centered columns (7 columns)
\hline %inserts double horizontal lines
\hline %inserts double horizontal lines
\\
Parameters&~~~GRO J1655-40 &~~~XTE J1859+226 & ~~~H1743-322&~~~XTE J1550-564&~~~GRS 1915+105 \\ [0.5ex]
\hline %inserts double horizontal lines
\hline %inserts double horizontal lines
\\
$M(M_{\bigodot})$& ~~ $5.8387^{+0.0393}_{-0.0390}$&~~~~$9.0156^{+0.1807}_{-0.1764}$&~~~~$10.8275^{+0.2756}_{-0.2719}$&~~~~$10.2178^{+0.2519}_{-0.2445}$        &~~~~$13.8884^{+0.4589}_{-0.4533}$ \\
\\
$a/M$&        ~~ $0.29^{+0.002}_{-0.002}$&~~~~$0.15^{+0.002}_{-0.002}$&~~~~$0.29^{+0.007}_{-0.007}$&~~~~$0.34^{+0.007}_{-0.007}$        &~~~~$0.28^{+0.015}_{-0.015}$ \\
\\
$r/M$&             ~~ $5.7992^{+0.0208}_{-0.0210}$&~~~~$6.8492^{+0.0665}_{-0.0691}$&~~~~$5.7460^{+0.0853}_{-0.0842}$ &~~~~$5.4771^{+0.0823}_{-0.0842}$         &~~~~$6.1501^{+0.1337}_{-0.1285}$ \\
\\
$|Q|/M$&            ~~ $< 0.16941$&~~~~$< 0.23632$&~~~~$< 0.30055$ &~~~~$< 0.26862$        &~~~~$< 0.24864$ \\
\\
$ \lambda/M^{4}$&       ~~ $< 0.24811$ &~~~~$< 0.77876$ &~~~~$<0.79043$ &~~~~$< 0.71135$        &~~~~$< 0.72226$ \\
\hline %inserts double horizontal lines
\hline %inserts double horizontal lines
\label{table:Best-fit}
\end{tabular}
\end{table}

%%%%%%%%%%%%%%%%%%%%%%%%%%%%%%%%%%%%%%%%%%%%%%%%%%%%%%%%%%%%%%%%%%%%%
Our results highlight the stringent effects of the magnetic charge and show that the constraints on the nonminimal parameter are below unity. In particular, we obtain the best-fit upper bounds on the magnetic charge and the nonminimal parameter as
\begin{equation}
|Q| < 0.16941 M,~~~~~~\lambda < 0.24M^{4}
\end{equation}
at the $90\%$ C.L.\ for $GRO~J1655-40$.
This comprehensive set of measured QPO observations reflects relatively small statistical uncertainties, and such rich data satisfy the requirements of the RPM, thereby enabling robust parameter estimation within the paradigm of rotating regular magnetic BH spacetime.
As a consequence, our MCMC analysis of $GRO~J1655-40$ is particularly suitable for probing possible deviations from the standard Kerr spacetime.

It is worth noting that QPOs have also been widely employed to constrain model parameters or to identify preferred spacetime geometries in various classes of regular BHs. For instance, a class of regular BHs arising from nonlinear electrodynamics (NED) was investigated, where the RPM of QPOs was used to constrain the NED charge parameters~\cite{Banerjee:2022chn,Hazarika:2025axz}. In addition, QPO analyses have been applied to place bounds on quantum gravity effects by modeling regular self-dual BHs~\cite{Liu:2023vfh,Wu:2025xtn}.
Moreover, the authors of Ref.~\cite{Jiang:2021ajk} utilized QPO observations of GRO J1655$-$40 to constrain the rotating Simpson–Visser spacetime, finding that the observational data favor a regular BH configuration with a single event horizon. In Ref.~\cite{Boshkayev:2023rhr}, various regular BH solutions were examined as effective exterior geometries for neutron stars using several QPO data sets. It was shown that regular BH spacetimes can successfully describe neutron star exteriors and, in many cases, provide a better fit to the data than the standard Schwarzschild geometry with spherical symmetry.
Taken together with our results, these studies demonstrate that X-ray observations of QPOs provide a powerful and complementary probe for testing gravity theories and constraining regular BH models when such objects are assumed to be the central compact sources.

In the next section, we examine the spin precession of a test gyroscope in the rotating regular magnetic BH spacetime. Note that in principle the following theoretical studies should focus on the parameter space constrained in previous section, but in order to better exhibit the phenomenal difference, we shall relax the parameters in the general regime.

\section{Spin precession of a test gyroscope}\label{sec:Spin Precession}

In this section, we analyze a test gyroscope to determine its spin precession frequency as measured by a stationary observer in the spacetime described by the metric given in Eq.~\eqref{Eq: metric}. Without loss of generality, we consider a stationary test gyroscope, labeled as a fixed-position gyroscope, for which the radial coordinate $r$ and the polar angle $\theta$ remain fixed outside the ergoregion, and we also examine the asymptotic behavior in the limit of spatial infinity.

In this setup for a fixed-position gyroscope, the four-velocity is represented as \cite{Chakraborty:2016mhx}
\begin{equation}
u^{\mu}_{\text{stationary}}= u^t_{\text{stationary}} (1, 0, 0, \Omega),
\end{equation}
where $t$ denotes the time coordinate and $\Omega=d\phi/dt$ represents the angular velocity of the observer.
The spin vector of a stationary gyroscope aligned with a timelike Killing vector field is Fermi–Walker transported along its integral curves
\begin{equation}
u=(-\mathbf{K}^{2})^{-1/2}\mathbf{K}, \label{Eq:Fermi-Walker}
\end{equation}
where $\mathbf{K}$ is the timelike Killing vector field.

Since the regular magnetic BH spacetime admits two Killing vectors, i.e., the timelike Killing vector $\partial_{t}$ and the azimuthal Killing vector $\partial_{\phi}$, we are therefore able to construct a more general Killing vector field $\mathbf{K}$ as
\begin{equation}
\mathbf{K} = \partial_{t} +\Omega \partial_{\phi}. \label{Eq:killing vector}
\end{equation}

In this scenario, the spin precession frequency of a stationary test gyroscope is identified with the rescaled vorticity field of the observer congruence and is succinctly expressed as the rescaled vorticity one-form of the congruence~\cite{Chakraborty:2013naa,Chakraborty:2016mhx,Wu:2023wld,Iyer:2025ccd} as follows:
\begin{equation}
\tilde{\Omega}_p=\frac{1}{2\mathbf{K}^{2}}\ast(\tilde{\mathbf{K}}\wedge d\tilde{\mathbf{K}}), \label{Eq:general spin precession}
\end{equation}
here, $\tilde{\mathbf{K}}$ is the covector of $\mathbf{K}$, $\ast$ represents the Hodge dual, and $\wedge$ denotes the wedge product.
Augmenting the approach presented in \cite{Chakraborty:2016mhx, Wu:2023wld, Iyer:2025ccd, Zahra:2025fvq}, we identify the vector corresponding to the spin precession frequency as
\begin{equation}
\vec{\Omega}_{p} = \frac{\varepsilon_{ckl}}{2\sqrt{-g}\left(1 + 2\Omega\frac{g_{0c}}{g_{00}} + \Omega^2 \frac{g_{cc}}{g_{00}}\right)} \left[\left( g_{0c,k} - \frac{g_{0c}}{g_{00}} g_{00,k} \right)+ \Omega \left( g_{cc,k} - \frac{g_{cc}}{g_{00}} g_{00,k} \right)+ \Omega^2 \left( \frac{g_{0c}}{g_{00}} g_{cc,k} - \frac{g_{cc}}{g_{00}} g_{0c,k} \right)\right] \partial_l, \label{Eq: spin precession}
\end{equation}
here, $g$ is the determinant of the metric $g_{\mu\nu}$ with $\mu,\nu=0, 1, 2, 3$, and $\varepsilon_{ckl}$ with $c,k,l=1,2,3$ is the Levi-Civita symbol.

By inserting the rotating regular magnetic metric, we obtain a simplified expression for the general spin precession as
\begin{equation}
\vec{\Omega}_{p} = \frac{(A_1 \hat{r} + A_2 \hat{\theta})}{2\sqrt{-g} \left( 1 + 2\Omega \frac{g_{t\phi}}{g_{tt}} + \Omega^2 \frac{g_{\phi\phi}}{g_{tt}} \right)},\label{Eq:total spin precession}
\end{equation}
with
\begin{align}
A_1 &= -\sqrt{g_{rr}} \left[ \left(g_{t\phi,\theta} - \frac{g_{t\phi}}{g_{tt}} g_{tt,\theta}\right) + \Omega \left( g_{\phi\phi,\theta} - \frac{g_{\phi\phi}}{g_{tt}} g_{tt,\theta} \right) + \Omega^2 \left( \frac{g_{t\phi}}{g_{tt}} g_{\phi\phi,\theta} - \frac{g_{\phi\phi}}{g_{tt}} g_{t\phi,\theta} \right) \right],\label{Eq: reduced form}
\end{align}
\begin{align}
A_2 &= \sqrt{g_{\theta\theta}} \left[ \left(g_{t\phi,r} - \frac{g_{t\phi}}{g_{tt}} g_{tt,r}\right) + \Omega \left( g_{\phi\phi,r} - \frac{g_{\phi\phi}}{g_{tt}} g_{tt,r} \right) + \Omega^2 \left( \frac{g_{t\phi}}{g_{tt}} g_{\phi\phi,r} - \frac{g_{\phi\phi}}{g_{tt}} g_{t\phi,r} \right) \right],
\label{Eq: reduced form1}
\end{align}
these expressions hold for observers inside and outside the ergosphere within a limited range of $\Omega$.

As in the cases of \cite{Wu:2023wld,Zhen:2025nah}, the validity of the observation is restricted to a specific domain of $\Omega$, with fixed values of $r$ and $\theta$ as
\begin{equation}
\Omega_{-}(r, \theta) < \Omega(r, \theta) < \Omega_{+}(r, \theta),
\quad \text{with bound defined by} \quad
\Omega_{\pm} = \frac{-g_{t\phi} \pm \sqrt{g_{t\phi}^{2} - g_{\phi\phi} g_{tt}}}{g_{\phi\phi}}.\label{Eq:angu}
\end{equation}

Due to the gyroscope's persistent angular velocity, its general spin precession decomposes into two distinct components: the LT precession arising from frame-dragging by the rotating source and the geodetic precession resulting from spacetime curvature.
In the subsequent analysis, we investigate three configurations: the LT precession, the general spin precession, and the geodetic precession, to decipher their physical consequences within the framework of a regular rotating BH spacetime geometry.

\subsection{Lense-Thirring Precession}
In this analysis, we discuss how the LT precession of a test gyroscope arises due to the effect of the frame-dragging frequency $\Omega_{\mathrm{LT}}$ in a rotating regular magnetic BH spacetime. When the angular velocity $\Omega = 0$, the general spin precession frequency $\vec{\Omega}_p$ reduces to the LT precession frequency $\vec\Omega_{\mathrm{LT}}$, and the gyroscope is fixed to a static observer who remains at rest relative to infinity, a condition physically viable only outside the ergoregion~\cite{Chakraborty:2013naa}.
In this stationary spacetime setup, the observer's four-velocity is given by $u^\mu_{\text{static}} = u^t_{\text{static}}(1,0,0,0)$ and aligns with the timelike Killing vector $\mathbf{K} = \partial_t$.

This reduction has been extensively studied in various rotating spacetimes~\cite{Wu:2023wld,Zhen:2025nah,Wu:2025ccc,Wu:2025xtn,Iyer:2025ccd}, thereby providing a sensitive and key probe for distinguishing different spacetime geometries and revealing their unique gravitational properties near the ergoregion.

The vectorial and magnitude forms of LT precession frequency for the regular magnetic spacetime are given by
\begin{equation}
\vec{\Omega}_{\mathrm{LT}} = \frac{1}{2\sqrt{-g}}\left[- \sqrt{g_{rr}} \left( g_{t\phi,\theta} - \frac{g_{t\phi}}{g_{tt}} g_{tt,\theta} \right) \hat{r}
+ \sqrt{g_{\theta\theta}} \left( g_{t\phi,r} - \frac{g_{t\phi}}{g_{tt}} g_{tt,r} \right) \hat{\theta}\right], \label{Eq: vector LT}
\end{equation}
\begin{equation}
\Omega_{\mathrm{LT}} = \frac{1}{2\sqrt{-g}}\sqrt{g_{rr} \left( g_{t\phi,\theta} - \frac{g_{t\phi}}{g_{tt}} g_{tt,\theta} \right)^2+ g_{\theta\theta}
\left( g_{t\phi,r} - \frac{g_{t\phi}}{g_{tt}} g_{tt,r} \right)^2}. \label{Eq: Magni. LT}
\end{equation}
To observe the physical impact of the LT precession frequency, we investigate four distinct cases regarding the four parameters, each involving the variation of a single parameter, and examine how $\Omega_{\mathrm{LT}}$ depends on these parameters, along with its implications for frame-dragging, the underlying spacetime structure, and potential observational signatures.

%%%%%%%%%%%%%%%%%%%%%%%%%%%%%%%%%
\begin{figure}[ht!]
\centering
(a)\includegraphics[width=6.5cm]{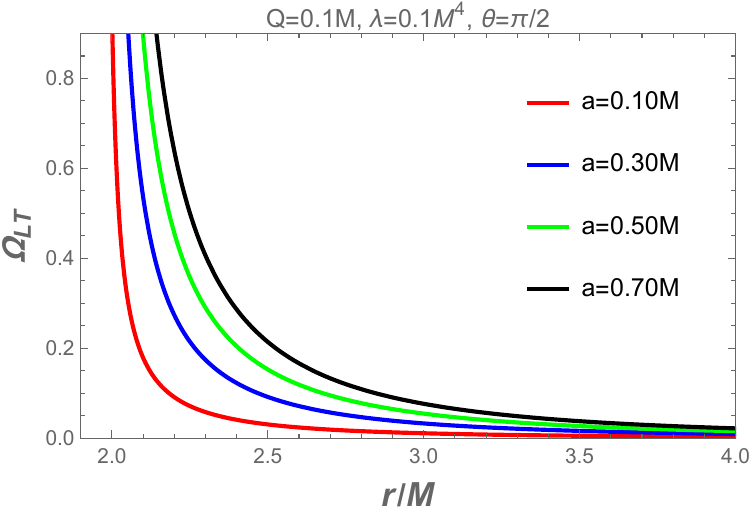}\hspace{1cm}
(b)\includegraphics[width=6.5cm]{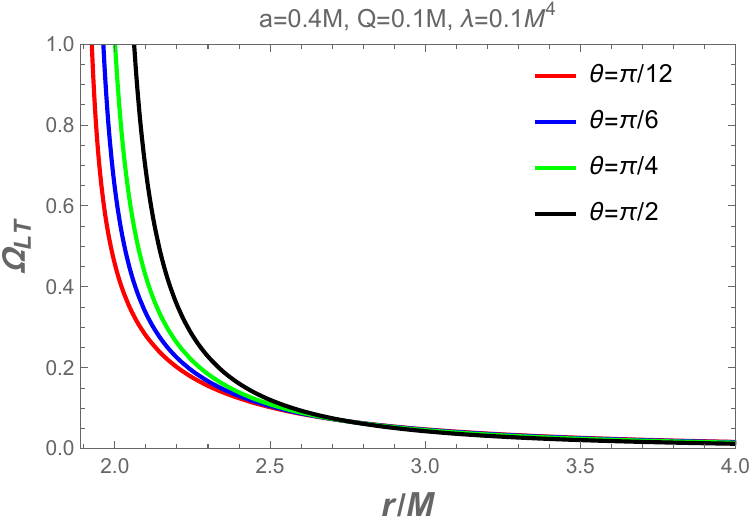}
(c)\includegraphics[width=6.5cm]{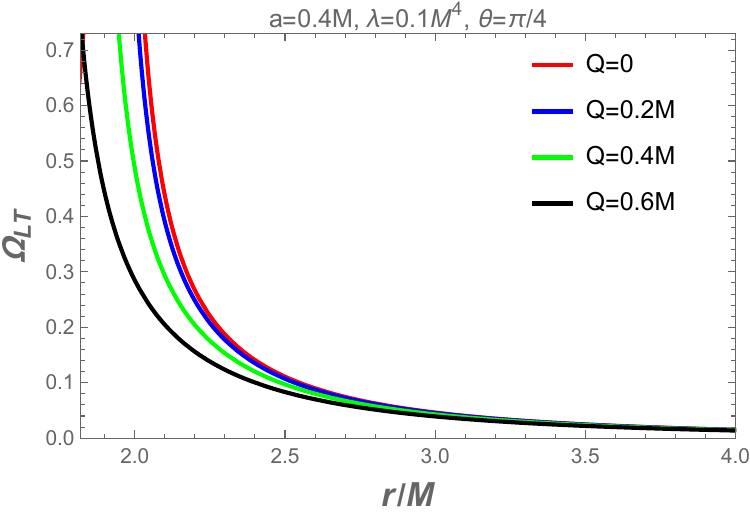}\hspace{1cm}
(d)\includegraphics[width=6.5cm]{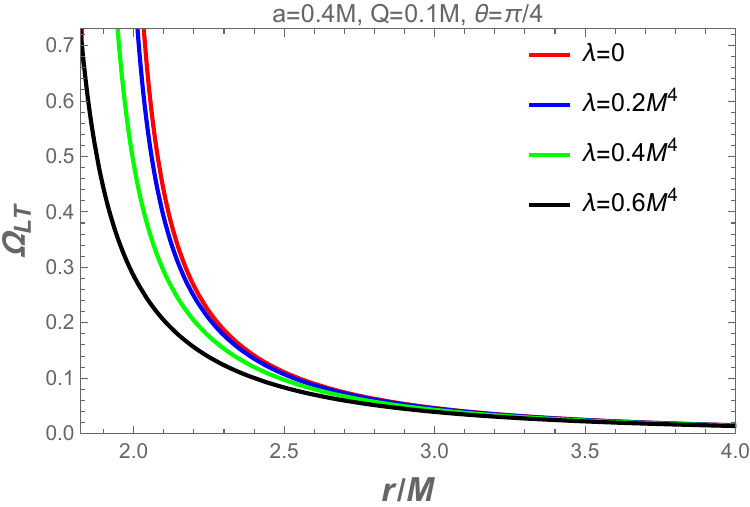}
\caption{
 The LT precession frequency is presented as a function of the radial coordinate for the regular magnetic BH. We observe the effects of the spin parameter in plot (a), with $Q/M = \lambda/M^{4} = 0.1$ and $\theta = \pi/2$; for plot (b), we observe the effects of the inclination angle with $Q/M = \lambda/M^{4} = 0.1$ and $a/M = 0.4$; in plot (c), we examine the effects of the magnetic charge with $\lambda/M^{4} = 0.1$, $a/M = 0.4$, and $\theta = \pi/4$; and for plot (d), we show the influence of the nonminimal parameter with $Q/M = 0.1$, $a/M = 0.4$, and $\theta = \pi/4$.}\label{fig:LT-overall}
\end{figure}

Fig.~\ref{fig:LT-overall} demonstrates the radial dependence of the LT precession frequency for the rotating regular magnetic BH, which diverges at the ergosphere.
Fig.~\ref{fig:LT-overall} (a) illustrates the dependence of $\Omega_{\mathrm{LT}}$ on the spin parameter $a/M$, showing that increasing spin strength acts as a source of frame-dragging, leading to stronger distortions of the local inertial frames near the ergosphere.
The LT precession frequency experiences strong frame-dragging effects as the polar angle moves toward the equatorial plane. In fact, this angular consistency is precisely what the Kerr metric predicts, and no deviation arises in its extensions, i.e., the hairy Kerr, Horndeski, and regular BH spacetimes. In our regular magnetic BH spacetime, a similar angular consistency holds.
No qualitative deviation from the Kerr spacetime~\cite{Chakraborty:2016mhx} or its extensions~\cite{Wu:2023wld,Zhen:2025nah,Wu:2025xtn} (e.g., hairy Kerr, Horndeski, or regular BHs) is observed.

Since the LT precession frequency essentially characterizes the rotation of the BH, such observations are reasonable, and both the BH spin and the gyroscope inclination angle are strongly interconnected.
Thus, the LT precession strengthens with increasing spin and polar angle, similar to the case of the Kerr BH spacetime~\cite{Chakraborty:2016mhx}.

In contrast, as the magnetic charge and nonminimal parameters of the regular BH increase, this does not imply an enhancement of the LT precession frequency $\Omega_{\mathrm{LT}}$, as shown in Fig.~\ref{fig:LT-overall} (c) and (d). In fact, as  the magnetic charge and the nonminimal parameters increase, the LT precession frequency profile is shifted toward the suppressed radial direction. This reduction in $\Omega_{\mathrm{LT}}$ indicates that the additional parameters of the regular magnetic BH spacetime dilute the frame-dragging effect.
%%%%%%%%%%%%%%%%%%%%%%%%%%%%%%%
%%%%%%%%%%%%%%%%%%%%%%%%%%%%%%%
\subsection{General spin precession frequency}
In this subsection, we probe the general spin precession frequency of a test gyroscope in the rotating regular magnetic BH spacetime.
We study the angular velocity constraints (Eq.~\eqref{Eq:angu}) for a gyroscope attached to a static observer. Therefore, we introduce a parameter $0<k<1$, such that the angular velocity $\Omega$ can be expressed in terms of the restricted range $\Omega_{\pm}$ as
\begin{equation}
\Omega = k \, \Omega_{+} + (1 - k) \, \Omega_{-} = \frac{(2k - 1) \sqrt{g_{t\phi}^2 - g_{tt} g_{\phi\phi}} - g_{t\phi}}{g_{\phi\phi}}.
\end{equation}
It is clear that for $k = 1/2$, this expression can be simplified as
\begin{equation}
\Omega \big|_{k=1/2} = - \frac{g_{t\phi}}{g_{\phi\phi}}.
\end{equation}
In this scenario, the stationary observer remains non-rotating with respect to the local spacetime frame, indicating that the observer possesses zero angular momentum and is therefore identified as a zero-angular-momentum observer (ZAMO) \cite{Bardeen:1972fi}.
In the Kerr case~\cite{Chakraborty:2016mhx}, the precession frequency of a gyroscope carried by a ZAMO exhibits distinct behavior, since the ZAMO gyroscope has no rotation relative to the local spacetime.

Substituting Eq.~\eqref{Eq:angu} into Eq.~\eqref{Eq:total spin precession}, we obtain a general expression for the spin precession frequency
\begin{equation}
\Omega_{p} = \frac{\sqrt{A_1^2 + A_2^2}}{2 \sqrt{-g} \left( 1 + 2\Omega \frac{g_{t\phi}}{g_{tt}} + \Omega^2 \frac{g_{\phi\phi}}{g_{tt}} \right)},
\end{equation}
where $A_1$ and $A_2$ are defined in Eqs.~\eqref{Eq: reduced form} and \eqref{Eq: reduced form1}.

%%%%%%%%%%%%%%%%%%%%%%%%%%%%%%%%%%%%%%%%%%%%%%%%%%%%%%%%%%%%%%%%%%%
\begin{figure}[ht!]
\centering
(a)\includegraphics[width=5cm]{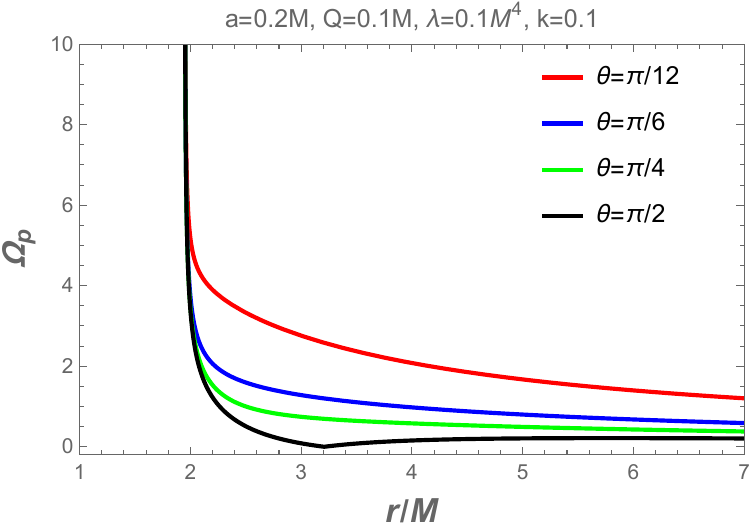}\hspace{1cm}
(b)\includegraphics[width=5cm]{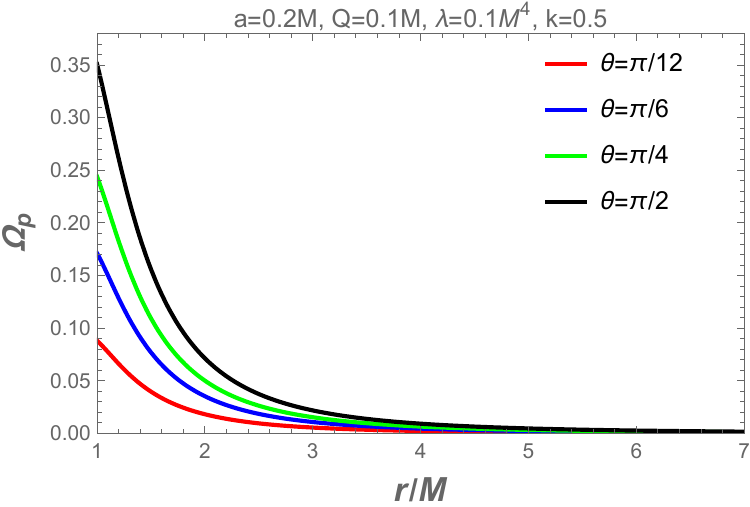}
(c)\includegraphics[width=5cm]{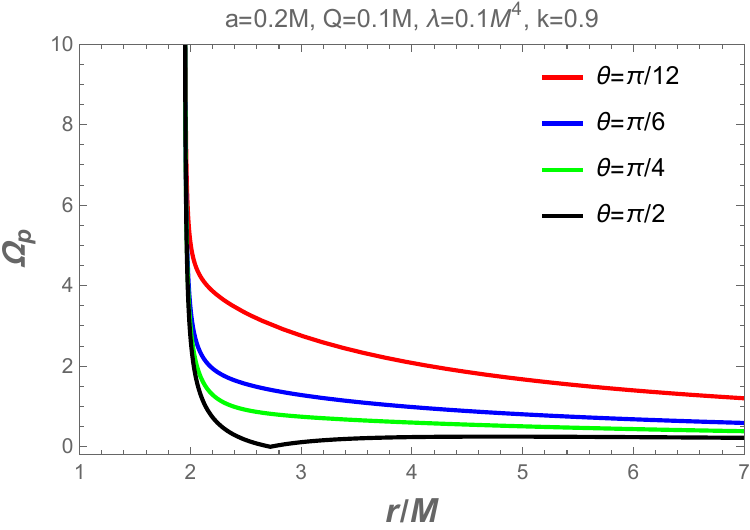}\hspace{1cm}
\caption{The effects of the spin precession as a function of the radial coordinate are illustrated, showing the influence of the inclination angle $\theta$ for different values of $k = 0.1, 0.5, 0.9$, with fixed parameters $Q/M = \lambda/M^4 = 0.1$ and $a/M = 0.2$, in the rotating regular magnetic BH spacetime.} \label{fig:GP-overall}
\end{figure}
Fig.~\ref{fig:GP-overall} shows the magnitude of the spin precession frequency for different values of ($k = 0.1, 0.5, 0.9$ from left to right) in the regular magnetic BH spacetime.
For $k = 0.1$ (Fig.~\ref{fig:GP-overall} (a)) and $k = 0.9$ (Fig.~\ref{fig:GP-overall} (c)), we observe that the spin precession frequency diverges near the BH horizon for all inclination angles.
It is noteworthy that, when $k = 0.5$ in Fig.~\ref{fig:GP-overall} (b), the observer coincides with the ZAMO.
Since the gyroscopes are effectively non-rotating relative to the local spacetime geometry and the stationary observers, the spin precession frequency remains finite for all directions.
%%%%%%%%%%%%%%%%%%%%%%%%%%%%%%%%%%%%%%%%%%%%%%%%%%%%%%%%%%%%%%%%%%%
\begin{figure}[ht!]
\centering
(a)\includegraphics[width=5.cm]{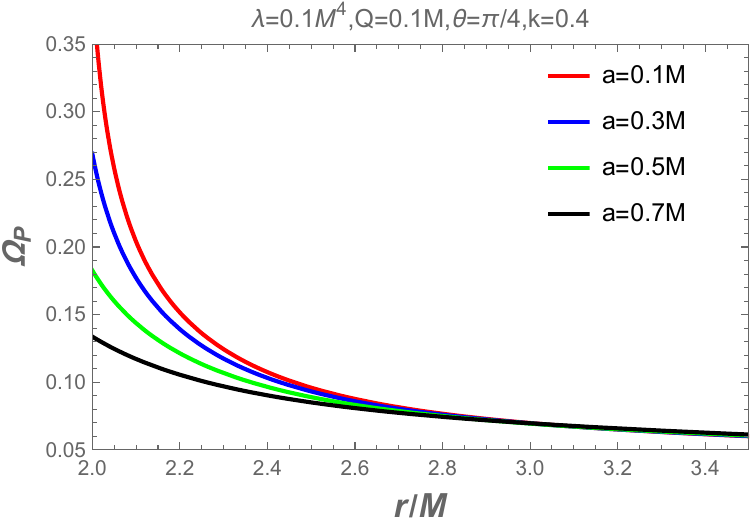}
(b)\includegraphics[width=5.cm]{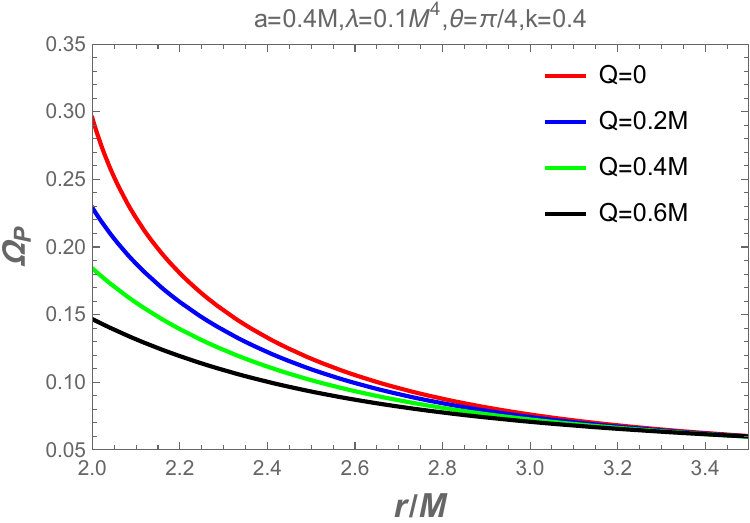}
(c)\includegraphics[width=5.cm]{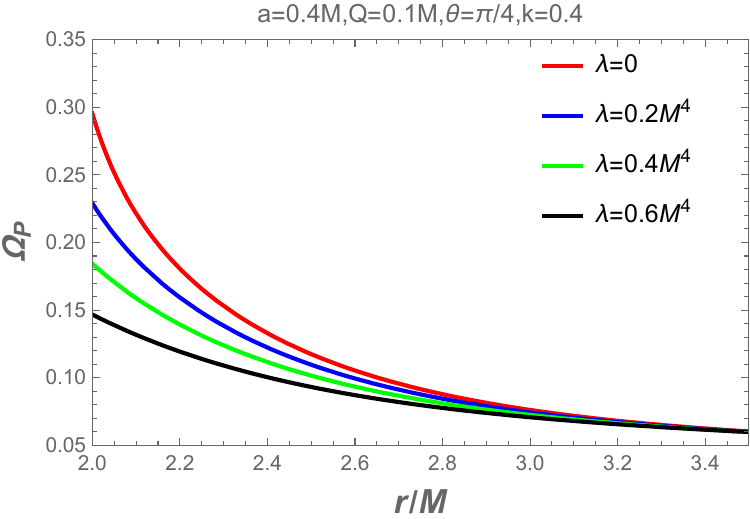}
\caption{ The spin precession is shown as a function of the radial coordinate for the regular magnetic BH. For plot (a), we analyze the effect of the spin parameter with $Q/M = 0.1$, $\lambda/M^{4} = 0.1$, $\theta = \pi/4$, and $k = 0.4$; for plot (b), we examine the influence of the magnetic charge $Q/M$ while keeping $\lambda/M^{4} = 0.1$, $a/M = 0.4$, $\theta = \pi/4$, and $k = 0.4$ fixed; for plot (c), we illustrate the effect of the nonminimal parameter with $Q/M = 0.1$, $a/M = 0.4$, $\theta = \pi/4$, and $k = 0.4$ held constant.}\label{fig:GP-parameters}
\end{figure}
The radial profile of the general spin precession frequency in the rotating regular magnetic BH spacetime, for different values of the model parameters, is also shown in Fig.~\ref{fig:GP-parameters}.
The figure reveals that, as the spin, magnetic charge, and nonminimal coupling parameters increase, the general spin precession frequency decreases monotonically, thereby suppressing the overall magnitude of the spin precession frequency.

%%%%%%%%%%%%%%%%%%%%%%%%%%%%%%%%%%%%%%%%%%%%%%%%%%%%%%%%%%%%%%%%%%%
\subsection{Geodetic precession frequency}
In the limiting case where $a=0$, the rotating regular BH described by the metric in Eq.~\eqref{Eq: metric} reduces to a static configuration~\cite{Balakin:2015gpq}, due to the absence of rotational angular momentum.
In this scenario, the LT precession frequency in Eq.~\eqref{Eq: Magni. LT} vanishes, while the spin precession frequency in Eq.~\eqref{Eq: spin precession} remains existent.
The geodetic precession frequency is a non-zero component contribution to the spin precession caused by spacetime curvature and takes the form:
\begin{equation}
\vec{\Omega}_{p}\Big|_{a=0} = \frac{1}{2\sqrt{-g}\left(1 + \Omega^2 \frac{g_{\phi\phi}}{g_{tt}}\right)} \left[-\Omega \sqrt{g_{rr}}\left(g_{\phi\phi,\theta} -
\frac{g_{\phi\phi}}{g_{tt}} g_{tt,\theta}\right) \hat{r} + \Omega \sqrt{g_{\theta\theta}}\left(g_{\phi\phi,r} - \frac{g_{\phi\phi}}{g_{tt}} g_{tt,r}\right) \hat{\theta}
\right],\tag{31}
\end{equation}
The spherical symmetry of the BH allows us to set $\theta = \pi/2$, thereby restricting the observers to the equatorial plane and simplifying the geodetic precession frequency as
\begin{equation}
\Omega_{p}\Big|_{a=0,\theta=\pi/2} = \frac{\Omega \sqrt{g_{\theta\theta}}\left(g_{\phi\phi,r} - \frac{g_{\phi\phi}}{g_{tt}} g_{tt,r}\right)}{2\sqrt{-g} \left(1
+ \Omega^2 \frac{g_{\phi\phi}}{g_{tt}}\right)}=\frac{r \sqrt{r^2} \Omega  \left(4 \lambda ^2+2 Q^2 r^6+r^8-3 r^7+4 \lambda  r^4+2 \lambda  r^3\right)}{\sqrt{r^4}
\left(2 \lambda +r^4\right) \left(2 \lambda +Q^2 r^2+r^6 \left(-\Omega ^2\right)+r^4-2 r^3-2 \lambda  r^2 \Omega ^2\right)}.
\end{equation}
Since a gyroscope moving in a static regular magnetic BH spacetime~\cite{Balakin:2015gpq} continues to experience precession, the angular frequency of precession coincides with the orbital angular velocity~\cite{Glendenning:1993di}, as
\begin{equation}
\Omega = \Omega_\phi = \left.\frac{d\phi}{dt}\right|_{a=0, \theta=\pi/2} =\frac{\sqrt{-Q^2 \left(r^4-2 \lambda \right)+r^5-6 \lambda  r}}{\left(2 \lambda +r^4\right)}.
\end{equation}
Thus, the expression for the geodetic precession can be obtained as follows:
\begin{equation}
\Omega_{p}\Big|_{a=0, \Omega=\Omega_\phi} = \frac{\sqrt{-Q^2 \left(r^4-2 \lambda \right)+r^5-6 \lambda  r}}{\left(2 \lambda +r^4\right)}.
\end{equation}
The above expression gives the precession frequency in Copernican coordinates, as measured with respect to proper time $\tau$.
To relate coordinate time $t$ and proper time $\tau$, we apply the transformation as
\begin{equation}
d\tau = \sqrt{\frac{4 \lambda ^2+2 Q^2 r^6+r^8-3 r^7+4 \lambda  r^4+2 \lambda  r^3}{\left(2 \lambda +r^4\right)^2}}dt.
\end{equation}
%%%%%%
\begin{figure}[ht!]
\centering
(a)\includegraphics[width=6.5cm]{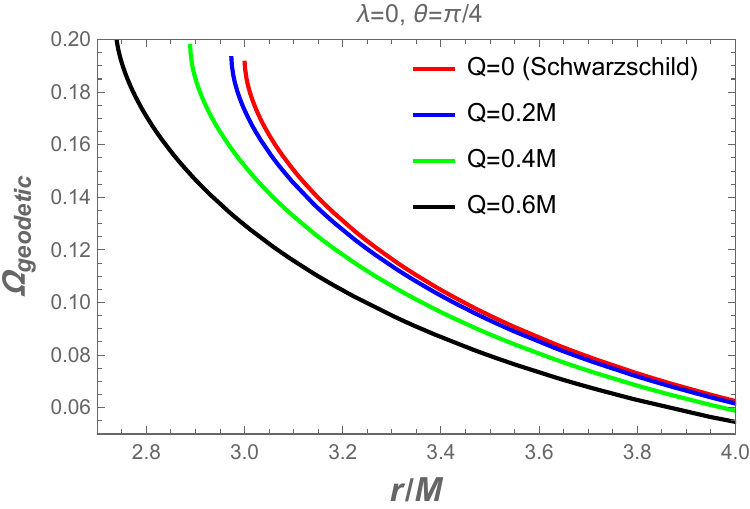}\hspace{1cm}
(b)\includegraphics[width=6.5cm]{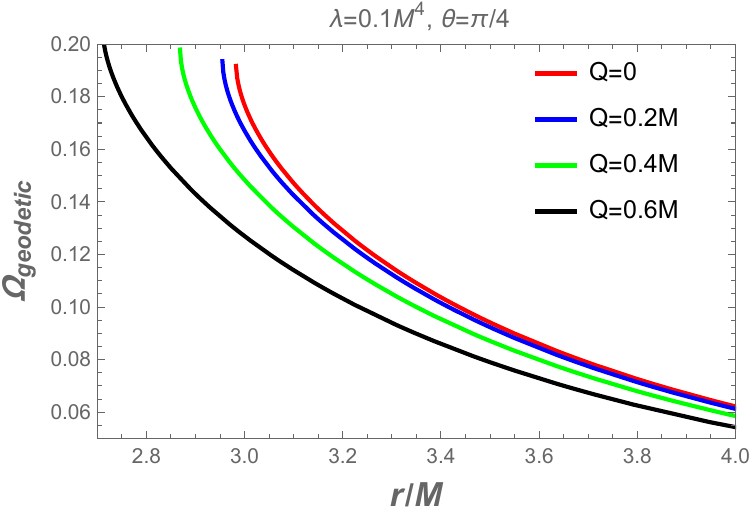}
(c)\includegraphics[width=6.5cm]{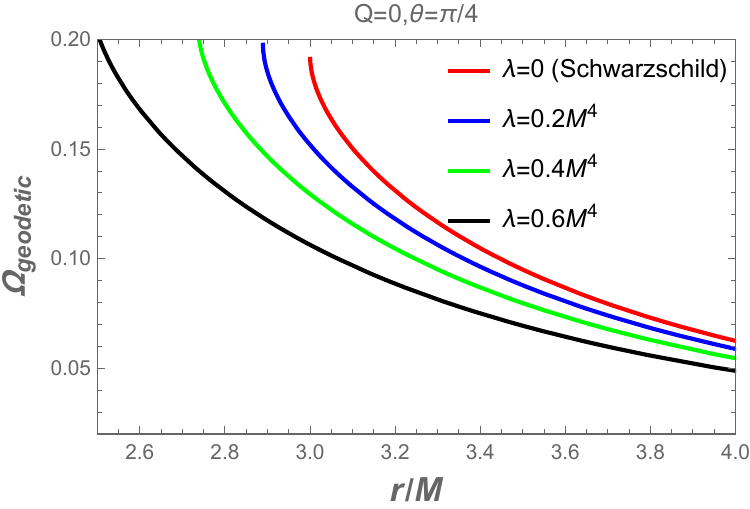}\hspace{1cm}
(d)\includegraphics[width=6.5cm]{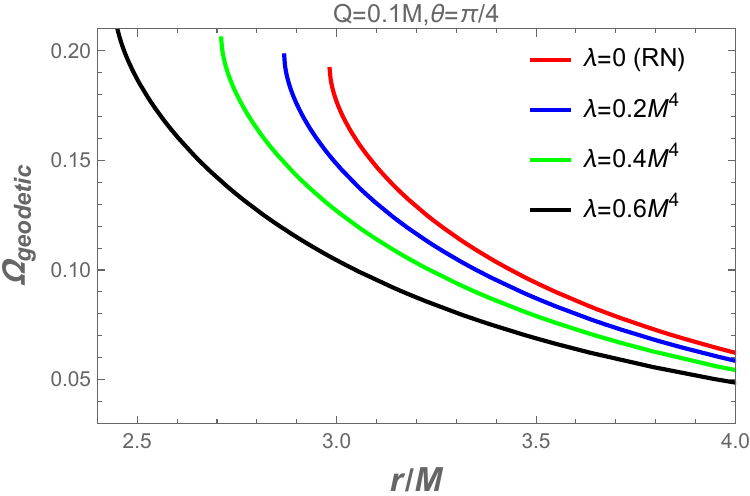}
\caption{The geodetic precession frequency is illustrated as a function of the radial coordinate in the static regular magnetic BH. The influence of $Q/M$ and $\lambda/M^{4}$ on the geodetic precession frequency is depicted for the static regular magnetic BH. In plots (a) and (b), we observe the effects of $Q/M$ with fixed $\lambda/M^{4} = 0, 0.1$ and $\theta = \pi/4$. In plots (c) and (d), we notice the effects of the nonminimal parameter $\lambda/M^{4}$ with fixed $Q/M = 0, 0.1$ and $\theta = \pi/4$.} \label{fig:geodetic}
\end{figure}
%%%%%%%
The geodetic precession frequency in coordinate time is expressed as
\begin{equation}
\Omega_{\text{geodetic}} =\sqrt{\frac{-Q^2 \left(r^4-2 \lambda \right)+r^5-6 \lambda  r}{\left(2 \lambda +r^4\right)^2}} \left(1-\sqrt{\frac{2 Q^2 r^6-3 r^7+2 \lambda  r^3}{\left(2 \lambda +r^4\right)^2}+1}\right).\label{Eq:geo}
\end{equation}
Since the geodetic precession derived from the spacetime curvature, it is evident that when the nonminimal coupling parameter, $\lambda/M^{4}$, vanishes, Eq.~\eqref{Eq:geo} reproduces the geodetic precession of the RN BH spacetime as reported in~\cite{Majumder:2025wsb}
\begin{equation}
\Omega_{\text{geodetic}}|_{\lambda=0} = \sqrt{\frac{r-Q^2}{r^4}} \left(1-\sqrt{\frac{2 Q^2+(r-3) r}{r^2}}\right).\label{Eq:geo-RN}
\end{equation}
Furthermore, when both $Q/M$, $\lambda/M^{4}$ vanish, the geodetic precession recovers the well-known result for the Schwarzschild BH~\cite{PhysRevD.19.2280,GyroscopeSchwarzschild1991}, thus confirming the validity of our derivation.
\begin{equation}
\Omega_{\text{geodetic}}|_{\lambda=Q=0} = \left(1-\sqrt{1-\frac{3}{r}}\right) \sqrt{\frac{1}{r^3}}.\label{Eq:geo-Sch}
\end{equation}

The physical effects of the model parameters $Q/M$ and $\lambda/M^{4}$ of the regular magnetic BH on the geodetic precession frequency are illustrated in Fig.~\ref{fig:geodetic}.
It is observed that the geodetic precession of the regular magnetic BH is suppressed with increasing values of both the magnetic charge parameter and the nonminimal parameter, and the deviation from the RN BH and the Schwarzschild BH becomes more pronounced, as shown in Fig.~\ref{fig:geodetic} (a) and (c). Similarly, the deviation of the geodetic precession frequency for the RN BH is less suppressed than that of the regular magnetic BH, as shown in Fig.~\ref{fig:geodetic} (d).

\section{Conclusion} \label{sec:conclusion}

In this paper, we investigate the orbital motion of a test particle, the frame-dragging effects, and the accretion disk physics associated with a rotating regular magnetic BH spacetime. We also examine the parameter constraints from the X-ray binary observations via QPOs and analyze the influence of the parameters on the spin precession of a test gyroscope in this rotating regular spacetime.
Our main findings and insightful results can be summarized as follows:

We computed the three fundamental frequencies, closely related to the QPOs in the regular magnetic spacetime, through small perturbations of the timelike bound circular orbit of a test particle on the equatorial plane.
We reveal that both the periastron and nodal precession frequencies exhibit monotonic suppression with increasing nonminimal parameters, consistent with the behavior in the Kerr-Newman and Kerr BH spacetimes. The spin parameter, on the other hand, enhances the nodal precession frequency while reducing the periastron precession frequency.

We also examine the dimensionless magnetic charge ($|Q|/M$) and the nonminimal coupling parameter ($\lambda/M^4$) to analyze the observational parameter constraints from QPO phenomena by performing an MCMC analysis using data from five X-ray binary sources ($GRO~J1655-40$, $XTE~J1859+226$, $H1743-322$, $XTE~J1550-564$, and $GRS~1915+105$).
Our joint analysis places stringent upper bounds on the magnetic charge ($|Q|/M$) and the nonminimal coupling parameter ($\lambda/M^4$) of the rotating regular magnetic BH.
Notably, the deviation parameters of all sources indicate that both $|Q|/M$ and $\lambda/M^4$ are less than 1, with the tight upper limit constraint observed from $GRO~J1655-40$. We find that the magnetic charge satisfies $|Q|/M < 0.16941$ and the nonminimal coupling parameter satisfies $\lambda/M^{4} < 0.24$ at the $90\%$ C.L.
Consequently, our observations provide a robust upper bound on the deviation parameters $|Q|/M$ and $\lambda/M^{4}$, while the best-fit values of the BH mass ($M~(M_{\bigodot})$), spin parameter ($a/M$), and the orbital radius ($r/M$) exhibit only minimal deviation predicted by the standard Kerr BH spacetime.
These findings demonstrate a novel mechanism for QPO generation beyond the conventional frame-dragging effect and indicate that QPO observations are in strong agreement with the Kerr spacetime, supporting the validity of GR in the strong gravity regime.

Furthermore, to investigate the physical implications of the model parameters, we consider the general parameter regime to study the spin precession of a test gyroscope attached to timelike stationary observers. This analytical development includes the LT precession, general spin precession, and geodetic precession frequencies. From this investigation, we found that the non-zero values of both the magnetic charge and nonminimal parameters thoroughly reduce the magnitudes of all three precession frequencies. We demonstrate that the spin precession frequency remains finite in the horizon limit for a ZAMO ($k = 0.5$), while it becomes divergent near the BH horizon for all other cases and inclination angles.

Next-generation X-ray observatories such as eXTP~\cite{eXTP:2018anb} and Athena~\cite{Nandra:2013jka}, with their substantially improved timing precision, spectral resolution, and sensitivity, are expected to deliver QPO measurements of unprecedented quality. We believe that further using these observations will enable more stringent tests on the current BH parameters.
Moreover, it will be  valuable to extend current QPO frequency and spin-precession frameworks to more realistic and extreme astrophysical scenarios, such as accretion flows that are non-axisymmetric, geometrically tilted, or disk warping and precession where relativistic effects are most pronounced. Incorporating such complexities will allow us to assess the limitations of the proposed models beyond idealized assumptions and to explore potential observational signatures that may distinguish different gravity theories or regular BH candidates.

\section*{Acknowledgments}
This work is partly supported by Natural Science Foundation of China under Grants No.12375054 and 12405067.
Meng-He Wu is also sponsored by Natural Science Foundation of Sichuan (No. 2025ZNSFSC0876).
H.G. is supported by the Institute for Basic Science (Grant No. IBS-R018-Y1).

%%%%%%%%%%%%%%%%%%%%%%%%%%%%%%%%%%%%%%%%%%%%%%%%%%%%%%%%%%%%%%%%%%%%%%%%%%%%%%%%%%%%%%%%%%%%%%%%%%%%%%%%%%%%%%%%%%%%%%%%%%%%%%%%%%%%%%%%%%%%%%%%%%%
\bibliography{ref}

\bibliographystyle{utphys}
%\bibliography{ref}

\end{document}